\documentclass[twocolumn]{aastex701}

\usepackage{booktabs}
\usepackage{graphicx}
\usepackage{amsmath}

\begin{document}

   \title{Turbulent viscosity in line-driven stellar winds}

\correspondingauthor{Matías Montesinos} \author[0000-0001-9789-5098]{Matías Montesinos} \affiliation{Departamento de Física, Universidad Técnica Federico Santa María, Avenida España 1680, Valparaíso, Chile} \email[show]{matias.montesinosa@usm.cl}

\author{Michel Cur\'e}
\affiliation{Instituto de F\'isica y Astronom\'ia, Universidad de Valpara\'iso, Av. Gran Breta\~na 1111, Casilla 5030, Valpara\'iso, Chile}
\email[]{michel.cure@uv.cl}

\author{Ignacio Araya}
\affiliation{Centro Multidisciplinario de F\'isica, Vicerrector\'ia de Investigaci\'on, Universidad Mayor, 8580745 Santiago, Chile}
\email{}

\author{Roberto O. J. Venero}
\altaffiliation{Member of the Carrera del Investigador Cient\'ifico, CONICET, Argentina}
\affiliation{Departamento de Espectroscop\'ia, Facultad de Ciencias Astron\'omicas y Geof\'isicas, Universidad Nacional de La Plata, Paseo del Bosque S/N, B1900FWA La Plata, Buenos Aires, Argentina}
\affiliation{Instituto de Astrof\'isica de La Plata, CCT La Plata, CONICET--UNLP, Paseo del Bosque S/N, B1900FWA La Plata, Argentina}
\email{}

\begin{abstract} Radiation-driven winds from massive stars generate powerful outflows known to be intrinsically turbulent. Despite this, standard hydrodynamic models have historically restricted the application of macroscopic turbulent viscosity to rotationally supported disks, neglecting its dynamic role in the supersonic expansion regime of the wind. We evaluated the dynamical relevance of macroscopic turbulent momentum transport in accelerating line-driven stellar outflows of rotating B supergiants. We introduced a generalized strain-limited viscous prescription within a stable, time-dependent 1D m-CAK hydrodynamic framework using the \texttt{FARGO3D} code. We compared inviscid, constant-viscosity, and strain-limited models to analyze their effects on the radial velocity, density stratification, mass-loss rate, and angular momentum transport. The strain-limited prescription produces a radially structured viscosity that primarily acts as a broad radial braking term. With the viscosity bounded through a prescribed maximum mixing length, the viscous acceleration remains subdominant to the total radiative acceleration but is dynamically relevant to the advective acceleration. In the strongest-viscosity model, the terminal velocity decreased by $\sim 39\%$ and the mass-loss rate by $\sim 58\%$ relative to the inviscid model in the stationary state. The flow remains radiatively driven, and angular momentum transport remains dominated by advection, maintaining a quasi-conserving $v_\phi \propto r^{-1}$ scaling. Turbulent viscosity exerts a measurable dynamic effect on supersonic line-driven winds. When bounded by the local macroscopic strain, it modifies the radial momentum balance and the resulting mass-loss rate without forcing the wind into a disk-like regime of strong viscous angular momentum redistribution. 
\end{abstract}

\section{Introduction}

Massive stars ($M \gtrsim 8\,M_\odot$), particularly early B-type stars and supergiants, drive powerful continuous material outflows via radiation pressure acting on the metal spectral lines \citep{cak, Pauldrach+1986, Kudritzki2000, Puls2008}. These radiatively driven winds are intrinsically structured and turbulent, far from being smooth or stationary. The fundamental line-deshadowing instability (LDI) \citep{Owocki1984, Owocki1988} induces small-scale velocity fluctuations and shocks
\citep{dessart2005}. Independently, radiation-hydrodynamic
simulations of the sub-photospheric turbulent zone driven by the
iron opacity bump also reveal structured, turbulent atmospheres in O
stars \citep{Moens2025}, indicating that intrinsic turbulence arises
in these stars through more than one physical mechanism.

Observationally, the presence of discrete absorption components (DACs) and rotational modulation further points to highly structured dynamic environments \citep{Prinja1995, Fullerton1997}. Despite advances in global hydrodynamic models, accurately predicting the mass-loss rates ($\dot{M}$) of B supergiants remains challenging. Significant discrepancies persist between theoretical predictions, evolutionary model requirements, and observational derivations, particularly across temperature domains, such as the bistability jump, where ionization shifts occur (e.g., \citealt{Haucke2018, Krticka2021, Krticka2024}). Consequently, exploring how macroscopic momentum transport and turbulent viscosity dynamically affect radial acceleration and the resulting mass-loss rate is a fundamental theoretical necessity.

Historically, efforts to couple rotation with wind expansion have highlighted complex dynamics, ranging from wind-compressed equatorial regions \citep{Bjorkman1993} and their inhibition by non-radial line forces \citep{Owocki1996} to the dense disk outflows of rotating B[e] supergiants \citep{Pelupessy2000}. However, the formal treatment of macroscopic viscous stresses maintains a strict bifurcation between radially expanding winds and rotationally supported equatorial disks. In the viscous decretion disk (VDD) paradigm \citep{lee1991}, angular momentum transport in the subsonic regime is governed by the standard Shakura-Sunyaev $\alpha$-prescription \citep{Shakura1973, okazaki2001}. This formulation scales the kinematic viscosity with the vertical scale height ($H \approx c_s/\omega_K$, where $c_s$ is
the sound speed and $\omega_K$ the Keplerian angular velocity), intrinsically assuming a vertical hydrostatic equilibrium and subsonic turbulence. Applying this geometric equivalence directly to a supersonic line-driven wind is thus physically invalid. In an open, radially accelerating flow, rapid expansion breaks the hydrostatic equilibrium. Extrapolating the classical VDD prescription into this regime forces the turbulent mixing length to diverge unphysically as the angular velocity declines, or conversely, shuts down macroscopic transport entirely if truncated arbitrarily.

Other viscosity prescriptions for accretion disks around black
holes have likewise been built around a critical effective Reynolds
number for the onset of turbulence, applied to the X-ray binary
Cygnus X-1 \citep{deFreitasPacheco1976}, to accretion disks more
broadly \citep{Duschl+2000}, and to a circumnuclear disk feeding a
growing supermassive black-hole seed \citep{Montesinos2011}.
In these prescriptions, turbulence is driven by the disk's rotational
motion, with the radial inflow treated as a slow, secondary drift.

In the context of line-driven winds, previous attempts to incorporate viscosity
into radiation hydrodynamics have relied on ad-hoc parameterizations
altering the centrifugal support \citep{cure2022} or on numerical
damping to stabilize LDI \citep{dessart2005}. More recently, m-CAK
line driving has been combined with a viscosity-mimicking rotational
prescription to explore quasi-Keplerian equatorial outflows in Be stars,
providing a complementary parameterized link between radiatively driven
winds, and viscous disk dynamics \citep{Cure+2026}.

A rigorous local Navier-Stokes formulation of turbulent momentum transport in an expanding wind requires abandoning the strict reliance on rotational shear. The central question is how to formulate a finite, local, and physically bounded turbulent viscosity for an open, accelerating, and radiatively driven outflow. To address this, we introduce a generalized strain-limited kinematic prescription. 

The prescription defines the turbulent mixing length from the local
deformation of the flow while imposing a finite maximum kinematic scale,
and explicitly incorporates both the radial expansion and rotational
shear.

To evaluate this formulation dynamically, a robust numerical framework that is free of steady-state topological artifacts is required. We utilize our highly stable, time-dependent 1D implementation of the m-CAK theory in the \texttt{FARGO3D} code \citep{Montesinos+2026}. We applied strain-limited viscosity to a fiducial rotating B-supergiant model. This target was selected strictly as a controlled hydrodynamic laboratory rather than an exhaustive representation of a specific observed object. The adopted m-CAK line-force parameters, specifically the overall line-driving strength ($k=0.32$) and the ionization parameter ($\delta=0.35$), along with a moderate rotation rate, establish the flow within the so-called $\delta$-slow hydrodynamic regime \citep{cure2011, Venero2024, Machuca2026}, providing a dense and stable baseline wind to isolate and measure the impact of macroscopic viscous stresses on the momentum balance.

The primary objective of this study is to evaluate whether a physically bounded turbulent viscosity can dynamically modify the radial expansion and mass-loss rate of a supersonic wind without unphysically driving strong disk-like angular momentum redistribution. To demonstrate how this novel formulation extends beyond the classical Shakura-Sunyaev paradigm, we contrast the wind dynamics generated by our strain-limited $\alpha_{\rm vis}$ prescription against standard inviscid and spatially uniform constant-viscosity baseline models. As this is a foundational methodological framework, the direct modeling of multidimensional LDI-induced clumping, non-isothermal viscous dissipation, complex asymmetric geometries, and object-specific synthetic observables is explicitly deferred to future investigations.


\section{Physical Model and Numerical Methods}

\subsection{Hydrodynamic equations}
\label{sec:hydrodynamic_equations}

We solve the time-dependent hydrodynamic equations on a
one-dimensional spherical radial grid in coordinates
$(r,\theta,\phi)$, where $r$ is the radius, $\theta$ the colatitude,
and $\phi$ the azimuth. We assume axisymmetry,
$\partial/\partial\phi=0$, no meridional motion, $v_\theta=0$, and no
meridional gradients, $\partial/\partial\theta=0$. All hydrodynamic
quantities therefore depend only on radius and time, with ${\bf v}=v_r\hat{\bf e}_r+v_\phi\hat{\bf e}_\phi$, where
$v_r$ and $v_\phi$ are the radial and azimuthal velocity components.

The evolved density variable is the volume density $\rho(r,t)$. No meridional structure or hydrostatic stratification is resolved in
the present one-dimensional framework.

The gas is assumed to be isothermal, with pressure
$P=\rho c_s^2$ and constant isothermal velocity scale
$c_s=(2k_B T_{\rm eff}/m_p)^{1/2}$, where $k_B$ is the
Boltzmann constant, $T_{\rm eff}$ the stellar effective temperature,
and $m_p$ the proton mass, following the implementation of
\citet{Montesinos+2026}.

Mass conservation in spherical coordinates is
\begin{equation}
\frac{\partial\rho}{\partial t}
+
\frac{1}{r^2}
\frac{\partial}{\partial r}
\left(
r^2\rho v_r
\right)
=
0 .
\label{eq:continuity_sph}
\end{equation}

The radial mass-loss rate is therefore
\begin{equation}
\dot{M}(r,t)
=
4\pi r^2\rho v_r .
\label{eq:mass_loss_rate}
\end{equation}
For a stationary solution, $\dot{M}$ is independent of radius.

The radial momentum equation is
\begin{equation}
\begin{aligned}
\frac{\partial v_r}{\partial t}
+
v_r\frac{\partial v_r}{\partial r}
-
\frac{v_\phi^2}{r}
={}&
-\frac{1}{\rho}\frac{\partial P}{\partial r}
-\frac{GM_\star(1-\Gamma_e)}{r^2}
\\
&+
g_{\rm line}
+
a_{{\rm visc},r},
\end{aligned}
\label{eq:radial_momentum_sph}
\end{equation}
where $G$ is the gravitational constant, $M_\star$ is the
stellar mass, $g_{\rm line}$ is the radiative
line acceleration, and $a_{{\rm visc},r}$ the radial viscous
acceleration. The electron-scattering Eddington factor is
$\Gamma_e=\sigma_eL_\star/(4\pi cGM_\star)$, with $\sigma_e$ the
electron-scattering opacity, $L_\star$ the stellar luminosity, and $c$
the speed of light, so the gravitational term
is written in the usual m-CAK reduced-gravity form.

In the numerical force decomposition used in \texttt{FARGO3D}, this is
equivalently represented by the pure Newtonian gravitational
acceleration, together with continuum and line radiative accelerations.
The line acceleration is described in Sect.~\ref{sec:radiative_line}.

The azimuthal momentum equation is
\begin{equation}
\frac{\partial v_\phi}{\partial t}
+
v_r\frac{\partial v_\phi}{\partial r}
+
\frac{v_rv_\phi}{r}
=
a_{{\rm visc},\phi},
\label{eq:azimuthal_momentum_sph}
\end{equation}
where $a_{{\rm visc},\phi}$ is the azimuthal viscous acceleration.
In the inviscid limit, this equation gives conservation of the specific
angular momentum $j=rv_\phi$.

We describe viscous momentum transport using the Newtonian
stress tensor together with the Stokes hypothesis, corresponding to
zero bulk viscosity \citep{Landau1987}. Defining the dynamic-viscosity coefficient as
$\eta=\rho\nu$, where $\nu$ is the kinematic viscosity, the stress
tensor is
\begin{equation}
{\boldsymbol{\mathcal T}}
=
2\eta
\left[
{\bf S}
-
\frac{1}{3}
\left(
\nabla\cdot{\bf v}
\right)
{\bf I}
\right],
\label{eq:visc_tensor_general}
\end{equation}
where ${\bf I}$ is the identity tensor and
${\bf S}=[\nabla{\bf v}+(\nabla{\bf v})^{\rm T}]/2$ is the
rate-of-strain tensor. For the present one-dimensional spherical flow,
\[
\nabla\cdot{\bf v}
=
\frac{\partial v_r}{\partial r}
+
\frac{2v_r}{r}.
\]

The nonzero stress components needed for the radial and azimuthal
momentum equations are
\begin{equation}
\mathcal{T}_{rr}
=
2\rho\nu
\frac{\partial v_r}{\partial r}
-
\frac{2}{3}\rho\nu
\left(
\nabla\cdot{\bf v}
\right),
\label{eq:tau_rr}
\end{equation}

\begin{equation}
\mathcal{T}_{\theta\theta}
=
\mathcal{T}_{\phi\phi}
=
2\rho\nu
\frac{v_r}{r}
-
\frac{2}{3}\rho\nu
\left(
\nabla\cdot{\bf v}
\right),
\label{eq:tau_thetatheta}
\end{equation}
and
\begin{equation}
\mathcal{T}_{r\phi}
=
\mathcal{T}_{\phi r}
=
\rho\nu r
\frac{\partial}{\partial r}
\left(
\frac{v_\phi}{r}
\right).
\label{eq:tau_rphi}
\end{equation}

The radial viscous acceleration is obtained from the spherical
divergence of the stress tensor,
\begin{equation}
a_{{\rm visc},r}
=
\frac{1}{\rho}
\left[
\frac{1}{r^2}
\frac{\partial}{\partial r}
\left(
r^2\mathcal{T}_{rr}
\right)
-
\frac{\mathcal{T}_{\theta\theta}
+
\mathcal{T}_{\phi\phi}}{r}
\right],
\label{eq:avisc_r_sph}
\end{equation}
whereas the azimuthal viscous acceleration is
\begin{equation}
a_{{\rm visc},\phi}
=
\frac{1}{\rho r^3}
\frac{\partial}{\partial r}
\left(
r^3\mathcal{T}_{r\phi}
\right).
\label{eq:avisc_phi_sph}
\end{equation}

The radial component $a_{{\rm visc},r}$ modifies the radial momentum
balance through the normal viscous stresses, whereas
$a_{{\rm visc},\phi}$ describes viscous angular-momentum transport
associated with the radial gradient of the local angular velocity
$\omega=v_\phi/r$.

The prescription for the kinematic viscosity $\nu$ is introduced in
Sect.~\ref{sec:viscosity_prescriptions}. Because the viscous stresses
depend on the product $\rho\nu$, spatial variations in both the density
and the kinematic viscosity contribute to the viscous acceleration. In
the inviscid limit, $\nu=0$, the system reduces to the time-dependent
spherical implementation of the m-CAK wind equations described by
\citet{Montesinos+2026}.

\subsection{Radiative Line Force}
\label{sec:radiative_line}

The radiative line acceleration is implemented following the modified
Castor, Abbott, and Klein \citep[m-CAK;][]{cak, Pauldrach+1986} theory.
In the spherical formulation, the hydrodynamic equations evolve the
volume density $\rho(r,t)$, and all density-dependent quantities
entering the m-CAK force are evaluated directly from $\rho$.

The line acceleration per unit mass is written as
\begin{equation}
g_{\rm line}
=
\frac{G M_\star \Gamma_e}{r^2}
\, k \,
t^{-\alpha_{\rm cak}}
\left(
\frac{n_e}{W(r)}
\right)^\delta
f_{\rm{disk}}\!\!
\left(
r,v_r,\frac{\partial v_r}{\partial r}
\right),
\label{eq:line_force}
\end{equation}
where $\Gamma_e$ is the electron-scattering Eddington factor,
$k$ sets the overall line-driving strength, $\alpha_{\rm cak}$ is the
power-law index of the line-strength distribution, and $\delta$
accounts for the sensitivity of the force to changes in the local
ionization state. The factor $GM_\star\Gamma_e/r^2$ provides the
continuum electron-scattering normalization of the m-CAK line force.

The dimensionless Sobolev optical-depth parameter is
\begin{equation}
t=
\frac{\sigma_e\rho v_{\rm th}}
{\left|\partial v_r/\partial r\right|},
\end{equation}
where $v_{\rm th}$ is
the proton thermal speed. The electron number density $n_e$ entering
the ionization correction is likewise evaluated directly from $\rho$.
No geometric density conversion is required in the spherical
formulation.

The geometrical dilution factor is
$W(r)=\frac{1}{2}[1-\sqrt{1-(R_\star/r)^2}]$, with $R_\star$
the stellar radius, while
$f_{\rm disk}$ is the finite-disk correction factor
\citep{friend1986, Pauldrach+1986}, which accounts for the finite angular
extent of the stellar surface.

The explicit expansion of $f_{\rm disk}$ and the associated m-CAK
regularity conditions are omitted here for conciseness. A detailed
description of the m-CAK formalism is given by
\citet{Cure_review_2023}, while its stable time-dependent numerical
implementation in the present \texttt{FARGO3D} wind framework is
described by \citet{Montesinos+2026}.

\subsection{Viscosity Prescriptions for Expanding Winds}
\label{sec:viscosity_prescriptions}

\subsubsection{The Strain-Limited $\alpha$-Prescription}

Standard viscous disk models commonly adopt the
\citet{Shakura1973} prescription, in which the characteristic
turbulent transport scale is tied to the local rotational dynamics.
Such a prescription cannot be transferred directly to a radially
accelerating line-driven wind, where neither Keplerian rotation nor a
hydrostatic scale height provides the relevant macroscopic length and
time scales.

Following the standard mixing-length interpretation of turbulent
transport, $\nu_{\rm turb}\sim v_{\rm turb}l_{\rm turb}$, we construct
a local mixing length from the resolved deformation timescale while
taking the characteristic unresolved turbulent velocity to be bounded
by the gas sound speed. This bound is a
modeling assumption, not a general statement that turbulent motions in
hot-star atmospheres or winds are subsonic. Radiation-hydrodynamic
simulations show that turbulent velocities can exceed the gas sound
speed while remaining below the radiation sound speed
\citep{Jiang2015,Debnath2024}. Similar motions may also be relevant for
B supergiants, but they are not included in the present closure.

We therefore introduce a strain-limited kinematic viscosity of the form

\begin{equation}
\nu(r)
=
\alpha_{\rm vis}
\frac{c_s^2}{S_{\rm eff}(r)},
\label{eq:nu_alpha}
\end{equation}
where $\alpha_{\rm vis}$ is a dimensionless efficiency
parameter and the effective strain-limiting rate is defined as
\begin{equation}
S_{\rm eff}(r)
=
\left[
\left(
\frac{d v_r}{d r}
\right)^2
+
\left(
r\frac{d\omega}{d r}
\right)^2
+
\left(
\frac{c_s}{l_{\max}}
\right)^2
\right]^{1/2},
\label{eq:seff}
\end{equation} 
with $\omega=v_\phi/r$ the local angular velocity and $l_{\max}$
the maximum mixing length. The first
contribution measures the radial velocity gradient, the second the
rotational shear, and the third imposes a lower bound on the effective
strain-limiting rate.

Here, $S_{\rm eff}$ is a phenomenological scalar rate that acts as an
inverse local mixing timescale. Through the scaling
$l_{\rm eff}\sim c_s/S_{\rm eff}$, it limits the effective turbulent
mixing length by combining the resolved radial velocity gradient and
rotational shear with the cutoff $c_s/l_{\max}$.

This particular construction of $S_{\rm eff}$ defines the strain-limited
closure proposed in this study. It provides a local and bounded
viscosity prescription for an expanding wind.

We adopt $l_{\max}=0.5 ~ r$, corresponding to
$S_{\min}=c_s/l_{\max}=2c_s/r$. The quantity $l_{\max}$ is a kinematic
upper bound on the turbulent mixing length and is not related to a
disk scale height. In particular, it should not be interpreted as a
vertical or meridional scale, since no meridional structure or
hydrostatic stratification is modeled in the present
one-dimensional spherical framework. We adopt this value as a fiducial choice.

Since $S_{\rm eff}\geq c_s/l_{\max}$, the prescription implies
$\nu\leq\alpha_{\rm vis}c_s l_{\max}$. The cutoff therefore prevents
the turbulent diffusivity from becoming arbitrarily large when the
resolved velocity gradients are small. It does not impose a
lower bound on $\nu$; whether the resulting viscosity remains
dynamically relevant must instead be determined from the relaxed
hydrodynamic solutions.

We restrict $0\leq\alpha_{\rm vis}\leq1$, so that
$\alpha_{\rm vis}$ retains the interpretation of a dimensionless
turbulent-transport efficiency. Here, $\alpha_{\rm vis}=0$ corresponds to the inviscid limit, whereas $\alpha_{\rm vis}=1$ represents the largest transport
efficiency explored in this study.

Neither $l_{\max}$ nor $\alpha_{\rm vis}$ is assigned a
universal physical value in the present model. They are phenomenological
control parameters: $\alpha_{\rm vis}$ specifies the efficiency of
turbulent transport, while $l_{\max}$ limits the mixing length. The
sensitivity to the adopted mixing length is tested using
$l_{\max}/r=0.25$, $0.5$, and $1$ in
Appendix~\ref{app:lmax_sensitivity}. Existing studies of the
line-driven instability identify a source of wind fluctuations
\citep{Owocki1984,Owocki1988}, but do not provide a unique calibration
of these parameters for the present viscosity closure. In a physical
line-driven wind, the effective transport may also depend on radial
expansion and rotational shear; these processes are represented
phenomenologically through $S_{\rm eff}$ in the present one-dimensional
model.

For comparison with the standard Shakura--Sunyaev
$\alpha$-viscosity prescription \citep{Shakura1973}, we consider the
limit in which rotational shear dominates $S_{\rm eff}$. In this case,
$S_{\rm eff}\simeq|r\,d\omega/dr|$ and
$\nu_{\rm shear}\simeq\alpha_{\rm vis}c_s^2/
|r\,d\omega/dr|$. For Keplerian rotation,
$r\,d\omega_{\rm K}/dr=-3\omega_{\rm K}/2$, giving
$\nu_{\rm shear}=(2/3)\alpha_{\rm vis}c_s^2/\omega_{\rm K}$.
This is equivalent to the original Shakura--Sunyaev stress
normalization, $\boldsymbol{|\mathcal{T}_{r\phi}|}=\alpha P$, in the
shear-dominated limit. The commonly used disk relation
$\nu=\alpha c_s H=\alpha c_s^2/\omega_{\rm K}$, valid for a vertically
hydrostatic Keplerian disk with $H=c_s/\omega_{\rm K}$, differs only by
the conventional numerical factor absorbed into the definition of
$\alpha$.

More generally, however, $\alpha_{\rm vis}$ should not be identified
with a local Shakura--Sunyaev $\alpha$, since radial expansion and the
mixing-length cutoff also contribute to $S_{\rm eff}$.

This comparison is used only to show that the adopted prescription
recovers the familiar shear-dominated scaling in the appropriate
limit. The models studied here are not disk solutions: they describe
radially accelerating, non-Keplerian equatorial outflows in a
one-dimensional spherical geometry. Accordingly, the adopted strain-limited
prescription is constructed directly from the radial velocity gradient,
rotational shear, and the maximum kinematic mixing length.

\subsubsection{Constant-Viscosity Model}

To isolate the effect of the radial dependence introduced by the
strain-limited prescription, we also consider models with a spatially
uniform kinematic viscosity. These models are evolved with the
same spherical hydrodynamic equations as the strain-limited models; only
the radial dependence of $\nu$ is removed.

The constant-$\nu$ values were chosen to approximately match
the effective viscosities of the corresponding relaxed strain-limited
solutions in the wind-launching region.
The adopted values are
$\nu=3.\times10^{-3}$ and
$7.3\times10^{-3}$ in code units for
$\alpha_{\rm vis}=0.5$ and $1.0$, respectively.
The inviscid model, with $\nu=\alpha_{\rm vis}=0$, is included as the
common reference.

This calibration provides constant-viscosity comparison models with
characteristic diffusivities similar to those produced by the
strain-limited prescription near the wind base. The comparison
therefore isolates the importance of the subsequent radial evolution
of $\nu(r)$ rather than simply contrasting models with unrelated
viscosity amplitudes.

\subsection{Stellar Model, Initial Conditions, and Numerical Setup}
\label{sec:numerical_setup}

\begin{table}
\caption{Adopted stellar and line-force parameters.}
\label{tab:model_setup}
\centering
\begin{tabular}{lcc}
\hline\hline
Quantity & Symbol & Value \\
\hline
\multicolumn{3}{c}{Stellar model} \\
\hline
Effective temperature & $T_{\rm eff}$ & $19000~{\rm K}$ \\
Surface gravity & $\log g$ & $2.50$ \\
Stellar radius & $R_\star$ & $40~R_\odot$ \\
Stellar mass & $M_\star$ & $18.45~M_\odot$ \\
Luminosity & $\log(L_\star/L_\odot)$ & $5.27$ \\
Isothermal sound speed & $c_s$ & $17.7~{\rm km~s^{-1}}$ \\
Electron Eddington factor & $\Gamma_e$ & $0.244$ \\
Critical velocity & $v_{\rm crit}$ & $258~{\rm km~s^{-1}}$ \\
Rotation parameter & $\Omega$ & $0.50$ \\
Equatorial rotation speed & $v_{\rm rot}$ & $129~{\rm km~s^{-1}}$ \\
\hline
\multicolumn{3}{c}{Line-force parameters} \\
\hline
CAK exponent & $\alpha_{\rm cak}$ & $0.50$ \\
Force multiplier & $k$ & $0.32$ \\
Ionization parameter & $\delta$ & $0.35$ \\
\hline
\end{tabular}
\end{table}

For all simulations, we adopt the T19 B-supergiant model from
\citet{Montesinos+2026}. The adopted stellar and line-force parameters
are summarized in Table~\ref{tab:model_setup}. The stellar rotation
parameter is $\Omega=v_{\rm rot}/v_{\rm crit}$, with
$v_{\rm crit}=[GM_\star(1-\Gamma_e)/R_\star]^{1/2}$. We adopt
$\Omega=0.5$, corresponding to
$v_{\rm rot}=129~{\rm km~s^{-1}}$.

The line-force parameters are
$\alpha_{\rm cak}=0.5$, $k=0.32$, and $\delta=0.35$.
The present one-dimensional model represents the radial flow along the
stellar equatorial plane, where the rotational contribution is maximal.
For the adopted values of $\Omega=0.5$ and $\delta=0.35$, this
equatorial solution lies in the $\delta$-slow hydrodynamic regime \citep{cure2011}.
Further details of the stellar model and the reference time-dependent
m-CAK solution are given by \citet{Montesinos+2026}.

The hydrodynamic equations are solved with \texttt{FARGO3D}
\citep{Benitez-Llambay+2016}, an Eulerian, finite-difference code
that implements the FARGO orbital-advection algorithm of
\citet{Masset2000}. Its upwind scheme introduces resolution-dependent
numerical dissipation in both the inviscid and viscous models,
whereas the prescribed kinematic viscosity enters explicitly
through the viscous stress tensor. All models are compared at the
same resolution; we do not separately quantify the numerical
viscosity.

We adopt a logarithmic radial grid extending from
$R_{\rm in}=R_\star$ to $R_{\rm out}=150R_\star$, with
$N_r=1024$ active cells. Although the computational grid extends to $150R_\star$,
all quantitative analyses and radial profiles presented below
are restricted to $r\leq100R_\star$. In all models, the
flow becomes supersonic close to the stellar surface, so the
hyperbolic, advective part of the flow becomes causally disconnected
from the outer boundary after the sonic point is crossed. The viscous
stress, however, introduces a diffusive contribution to the radial
momentum equation that is not limited by the sonic transition and may
remain sensitive to the outer boundary. The interval
$100R_\star<r\leq150R_\star$ therefore serves as a numerical buffer,
reducing possible outer-boundary effects on the reported results.

In the active radial domain, the initial wind is specified by
\begin{equation}
\begin{aligned}
v_r(r,0)
&=
v_{\infty,0}
\left(
1-\frac{R_\star}{r}
\right)^\beta,
\\
\rho(r,0)
&=
\frac{\dot{M}_0}
{4\pi r^2 v_r(r,0)},
\\
v_\phi(r,0)
&=
v_{\rm rot}\frac{R_\star}{r}.
\end{aligned}
\label{eq:initial_conditions}
\end{equation}

The initial profiles are evaluated at radial cell centers
with $r>R_\star$ (the innermost at $r\simeq1.002R_\star$),
so $v_r(r,0)$ remains finite and positive, while the inner ghost cells
are set to the boundary state of Eq.~(\ref{eq:inner_boundary}).

We adopted $\beta=1$ for the initial radial velocity profile. Although the
adopted line-force parameters lead to a $\delta$-slow wind, this
$\beta$-law serves only as a smooth initial condition and does not
represent the relaxed $\delta$-slow solution. The azimuthal profile
initially conserves the specific angular momentum. The quantities
$v_{\infty,0}$ (initial terminal velocity) and $\dot{M}_0$
(initial mass-loss rate) specify only the initial guess.
Neither quantity fixes the final wind solution: the relaxed velocity,
density profile, and mass-loss rate emerge
from the subsequent time-dependent evolution.

The density in the inner ghost cells, $\rho_{\rm base}$,
and the radial velocity at the innermost radial interface,
$v_{\rm base}$, are fixed at
\begin{equation}
\begin{aligned}
\rho_{\rm base}
&=
10^{-11}\,{\rm g\,cm^{-3}},
\\
v_{\rm base}
&=
\frac{c_s}{4}.
\end{aligned}
\label{eq:inner_boundary}
\end{equation}

These conditions provide a fixed subsonic inner boundary from which the wind adjusts dynamically.

The simulations are performed in dimensionless code units, adopting
$\ell_0=1~{\rm AU}$ and $M_0=1~M_\odot$ as the units of length and mass,
respectively. The corresponding time unit is
$t_0=(\ell_0^3/G_{\rm cgs}M_0)^{1/2}$, with
$v_0=\ell_0/t_0$ and
$\rho_0=M_0/\ell_0^3$.
With this normalization, the gravitational constant is unity in code
units. All dimensional quantities reported below are converted back to
physical units using these scales.

\section{Results}

All simulations reached a stationary state after approximately
16 yr of evolution and remained stationary until the end of the
50 yr integrations. Unless explicitly showing the time evolution,
all profiles and quantitative diagnostics reported below correspond
to the fully relaxed winds at the final output, $t\simeq50$ yr.

\subsection{Global wind response to increasing $\alpha_{\rm vis}$}

\begin{figure}
\centering
\includegraphics[width=\hsize]{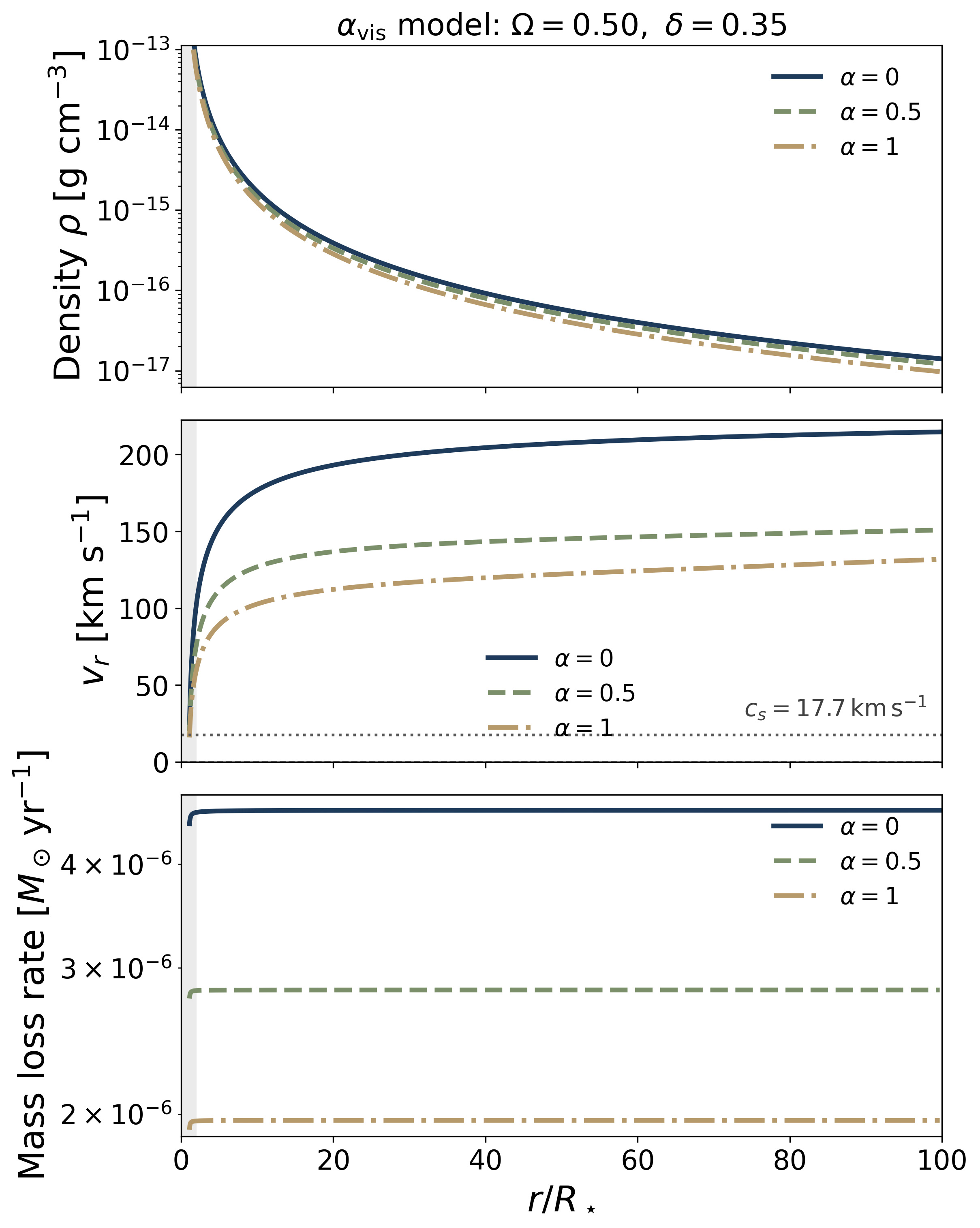}
\caption{
Response of the relaxed T19 wind structure to increasing
$\alpha_{\rm vis}$, for $\Omega=0.50$ and $\delta=0.35$.
Top: Volumetric density $\rho(r)$.
Middle: Radial velocity $v_r(r)$. The horizontal dotted line marks
the isothermal sound speed, $c_s=17.7~{\rm km~s^{-1}}$.
Bottom: mass-loss rate $\dot{M}(r)$.
Increasing $\alpha_{\rm vis}$ shifts the relaxed wind solution toward
lower density, radial velocity, and mass-loss rate.
The grey shaded region marks the inner boundary zone
excluded from quantitative analysis.}

\label{fig:alpha_vel}
\end{figure}

Figure~\ref{fig:alpha_vel} shows how the relaxed radial structure of
the T19 wind model with $\Omega=0.50$ and $\delta=0.35$ changes as
$\alpha_{\rm vis}$ increases. The inviscid case,
$\alpha_{\rm vis}=0$ provides the reference m-CAK solution. As
$\alpha_{\rm vis}$ is increased, the wind systematically relaxes to
solutions with lower radial velocity and lower volumetric
density throughout most of the domain.

The mass-loss profiles remain flat with radius,
consistent with stationary mass conservation.
However, their normalization decreases with increasing
viscosity. The characteristic mass-loss rate decreases from
$\dot{M}\simeq 4.64 \times10^{-6}\,M_\odot\,{\rm yr}^{-1}$
in the inviscid model to
$\simeq 2.82\times10^{-6}$ and
$\simeq 1.96\times10^{-6}\,M_\odot\,{\rm yr}^{-1}$
for $\alpha_{\rm vis}=0.5$ and $1.0$, respectively.

Thus, the strongest $\alpha_{\rm vis}$ model reaches a mass-loss rate
roughly a factor of 2.36 lower than the inviscid solution. The response
is not confined to the wind base: the radial velocity,
$\rho$, and mass-loss profiles are shifted over the full
relaxed flow. The dynamical origin of this global reduction is examined
below, through the radial viscous acceleration.

\subsection{Radial momentum balance and viscous braking}

\begin{figure}
\centering
\includegraphics[width=\hsize]{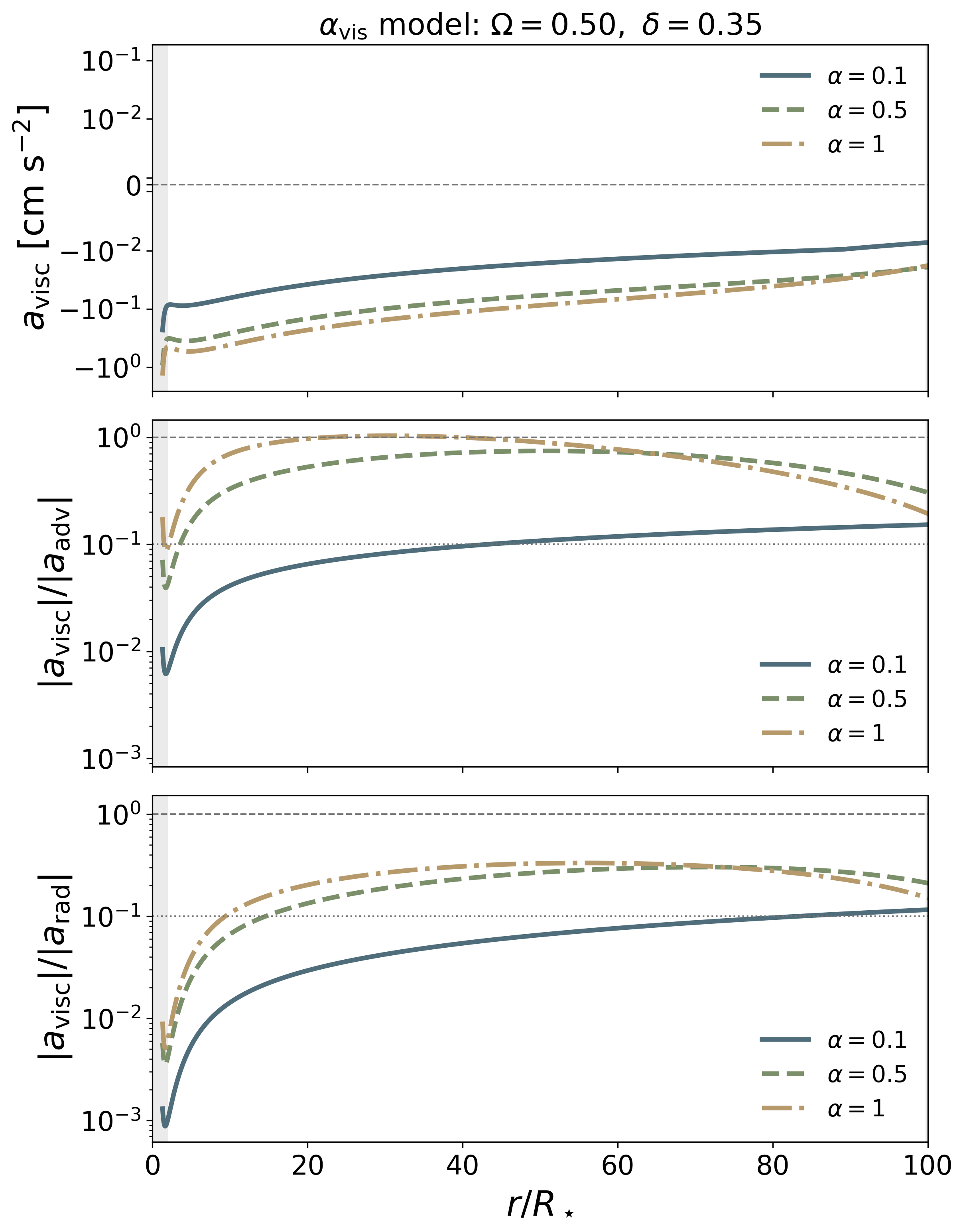}
\caption{
Radial viscous acceleration for the $\alpha_{\rm vis}>0$ models.
Top: Signed radial viscous acceleration $a_{{\rm visc},r}(r)$.
Middle: ratio between the radial viscous acceleration and the radial
advective acceleration,
$|a_{{\rm visc},r}|/|a_{\rm adv}|$.
Bottom: ratio between the radial viscous acceleration and the total
radiative acceleration,
$|a_{{\rm visc},r}|/|a_{\rm rad}|$.
The grey shaded regions represent the trimmed boundary zones that are excluded from quantitative analysis.
}
\label{fig:acc_alpha}
\end{figure}

Figure~\ref{fig:acc_alpha} shows the radial viscous acceleration for
models with a nonzero $\alpha_{\rm vis}$. The upper panel shows that
$a_{{\rm visc},r}$ remains negative throughout the radial domain for
all three models, with its magnitude increasing systematically with
$\alpha_{\rm vis}$. Viscosity therefore acts as a radial braking term
throughout the wind.

To quantify its dynamical relevance, we compare
$a_{{\rm visc},r}$ with the radial advective acceleration,
$a_{\rm adv}=v_r\,\partial v_r/\partial r$, and with the total
radiative acceleration,
$a_{\rm rad}=g_{\rm cont}+g_{\rm line}$, where
$g_{\rm cont}=GM_\star\Gamma_e/r^2$ is the continuum
electron scattering acceleration. In this decomposition, gravity is
therefore represented by the pure Newtonian term
$a_{\rm grav}=-GM_\star/r^2$. This is algebraically equivalent to the
reduced-gravity form of Eq.~(\ref{eq:radial_momentum_sph}), where the
continuum contribution is included through the factor $(1-\Gamma_e)$.

The median magnitude of the radial viscous acceleration increases from
$5.06\times10^{-2}\,{\rm cm\,s^{-2}}$ for
$\alpha_{\rm vis}=0.1$ to
$3.17\times10^{-1}\,{\rm cm\,s^{-2}}$ for
$\alpha_{\rm vis}=1.0$, a factor of approximately 6.3.

The middle panel shows that the viscous term becomes a significant
fraction of advective acceleration. The median ratio
$|a_{{\rm visc},r}|/|a_{\rm adv}|$ increases from
$5.26\times10^{-2}$ for $\alpha_{\rm vis}=0.1$ to
$6.72\times10^{-1}$ for $\alpha_{\rm vis}=1.0$,
corresponding to an increase of slightly more than one order
of magnitude. Thus, in the strongest-viscosity model,
the median ratio is approximately 2/3, with local
values reaching order unity, and the viscous term
can no longer be regarded as a negligible correction.

The lower panel of Fig.~\ref{fig:acc_alpha} shows that the viscous
acceleration remains subdominant to the total radiative acceleration.
The median ratio $|a_{{\rm visc},r}|/|a_{\rm rad}|$ increases from
$2.10\times10^{-2}$ for $\alpha_{\rm vis}=0.1$ to
$1.53\times10^{-1}$ for $\alpha_{\rm vis}=1.0$.
Even in the strongest-viscosity model, the median viscous acceleration
is therefore about $15\%$ of the total radiative acceleration.
The wind remains radiatively driven, while viscosity provides a
dynamically relevant negative contribution to the radial momentum
balance.

\subsection{Radial structure of the effective viscosity}

\begin{figure}
\centering
\includegraphics[width=\hsize]{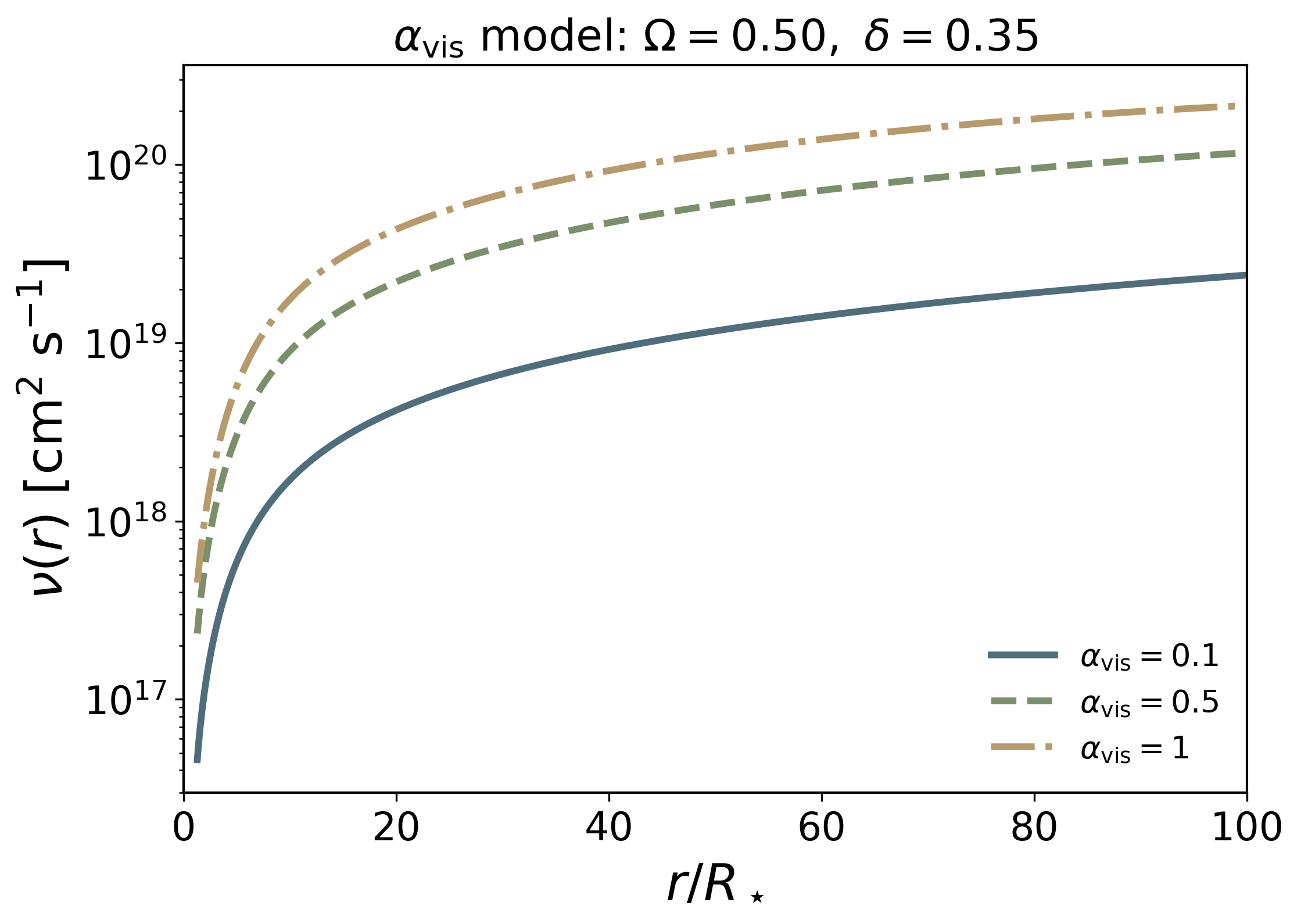}
\caption{
Radial profile of the kinematic viscosity generated by the
strain-limited $\alpha_{\rm vis}$ prescription for the relaxed T19
models with $\Omega=0.50$ and $\delta=0.35$.
The profiles correspond to $\alpha_{\rm vis}=0.1$, 0.5, and 1.0.
}
\label{fig:alpha_nu_profile}
\end{figure}

Figure~\ref{fig:alpha_nu_profile} shows the kinematic viscosity
profiles obtained from relaxed $\alpha_{\rm vis}$ models. Unlike
the constant-viscosity prescription, the strain-limited model produces
a radially structured viscosity. For all nonzero values of
$\alpha_{\rm vis}$, $\nu(r)$ rises rapidly from the wind base and
continues to increase throughout the plotted radial range.

This behavior directly follows from the adopted viscosity law. Since
the models are isothermal, $c_s$ is constant and the radial variation
of $\nu(r)$ is determined by the inverse of the effective strain-limiting
rate, $S_{\rm eff}^{-1}$. Near the wind base, the velocity gradients
entering $S_{\rm eff}$ are relatively large, keeping the kinematic
viscosity small. Farther out, these gradients generally decrease,
allowing the effective viscosity to increase.

The amplitude of the viscosity scales approximately with
$\alpha_{\rm vis}$, although the scaling is not purely multiplicative
because the relaxed velocity field, and therefore $S_{\rm eff}$,
changes with viscosity. The median kinematic viscosities are
$2.72\times10^{18}$,
$1.44\times10^{19}$, and
$2.83\times10^{19}\,{\rm cm^2\,s^{-1}}$
for $\alpha_{\rm vis}=0.1$, 0.5, and 1.0, respectively. Thus,
increasing $\alpha_{\rm vis}$ from 0.1 to 1.0 raises the characteristic
viscosity by approximately one order of magnitude.

The resulting profiles therefore differ fundamentally from a uniform
diffusion coefficient. At each radius, the local value of $\nu$ is
determined by the velocity gradients entering the strain-limited
prescription, together with the maximum mixing length cutoff. The
relative importance of these contributions is quantified below.

\subsection{Decomposition of the effective strain-limiting rate}
\label{subsec:seff_decomposition}

To determine which terms control the viscosity profiles shown in
Fig.~\ref{fig:alpha_nu_profile}, we decompose the effective
strain-limiting rate as 
\begin{equation}
S_{\rm eff}^2
=
S_r^2
+
S_\omega^2
+
S_{\min}^2,
\label{eq:seff_decomposition}
\end{equation}where $S_r=dv_r/dr$ is the radial velocity-gradient contribution,
$S_\omega=r\,d\omega/dr=dv_\phi/dr-v_\phi/r$ is the rotational-shear
contribution, and $S_{\min}=c_s/l_{\max}$ is the cutoff associated with
the maximum mixing length.

We emphasize that $S_r=dv_r/dr$ is the radial velocity-gradient term
adopted in the viscosity prescription; it should not be identified with
the full spherical velocity divergence,
$\nabla\cdot{\bf v}=dv_r/dr+2v_r/r$.
Thus, $S_{\rm eff}$ is the phenomenological scalar strain-limiting rate defining the present viscosity closure rather than the full invariant
of the spherical rate-of-strain tensor.

According to Eq.~(\ref{eq:nu_alpha}), the relative magnitudes of
$S_r$, $S_\omega$, and $S_{\min}$ determine whether the local viscosity
is controlled primarily by the radial velocity gradient, rotational
shear, or the maximum mixing length cutoff.

\begin{figure}
\centering
\includegraphics[width=\columnwidth]{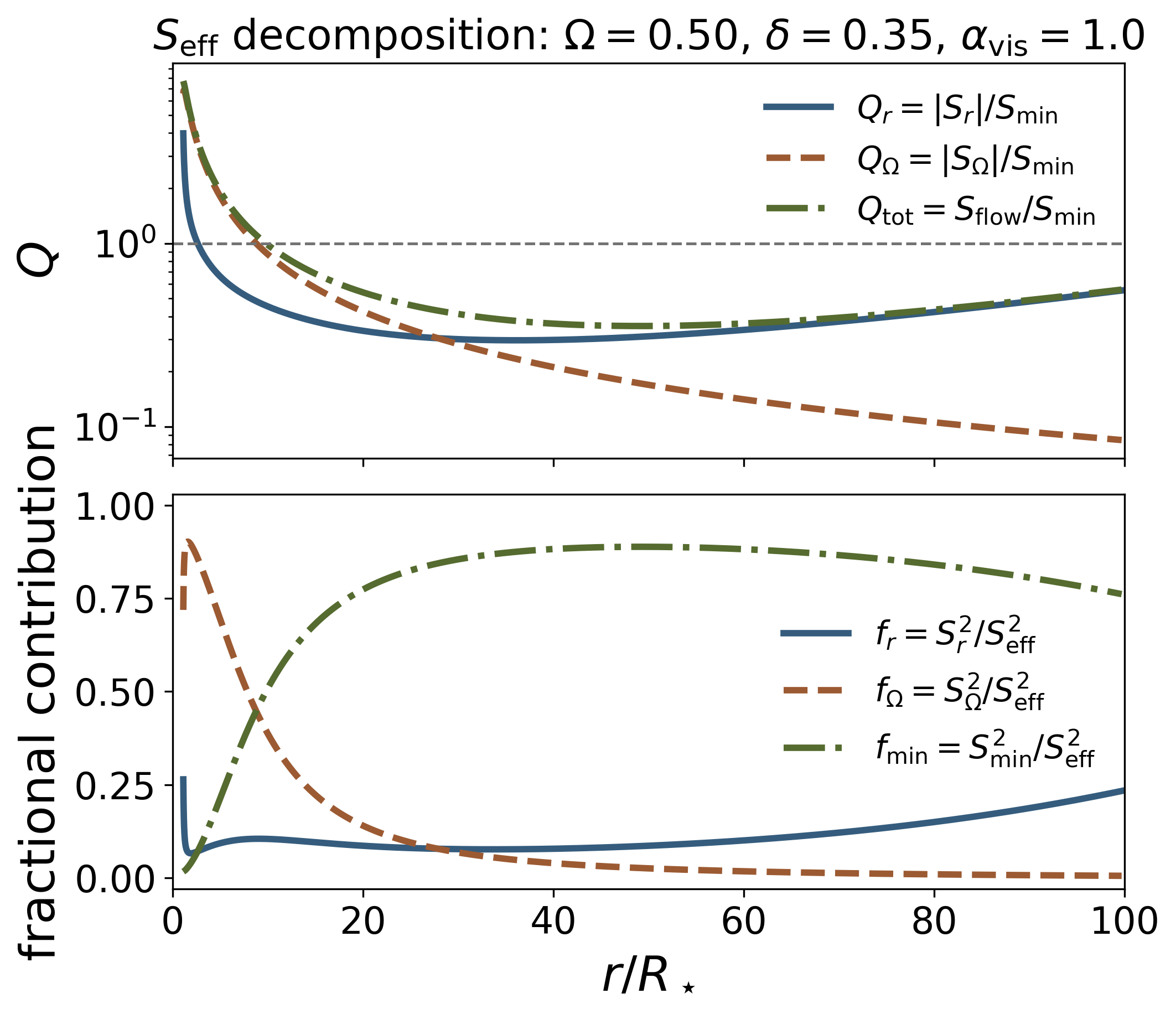}
\caption{
Decomposition of the effective strain-limiting rate for the strongest
strain-limited model, with $\Omega=0.50$, $\delta=0.35$, and
$\alpha_{\rm vis}=1$.
Top: ratios $Q_r=|S_r|/S_{\min}$,
$Q_\omega=|S_\omega|/S_{\min}$, and
$Q_{\rm tot}=\sqrt{S_r^2+S_\omega^2}/S_{\min}$.
Bottom: fractional contributions
$f_r=S_r^2/S_{\rm eff}^2$,
$f_\omega=S_\omega^2/S_{\rm eff}^2$, and
$f_{\min}=S_{\min}^2/S_{\rm eff}^2$.
The radial velocity-gradient term does not dominate the viscosity
denominator. Rotational shear provides the largest contribution in the
inner wind, whereas the cutoff controls much of the intermediate and
outer wind.
}
\label{fig:seff_decomposition}
\end{figure}

Figure~\ref{fig:seff_decomposition} shows this decomposition for the
strongest strain-limited model, with $\alpha_{\rm vis}=1$,
$\Omega=0.50$, and $\delta=0.35$.

To compare the amplitudes of the resolved velocity-gradient terms
with the cutoff, we define
\begin{equation}
Q_r
=
\frac{|S_r|}{S_{\min}},
\qquad
Q_\omega
=
\frac{|S_\omega|}{S_{\min}},
\qquad
Q_{\rm tot}
=
\frac{\sqrt{S_r^2+S_\omega^2}}{S_{\min}}.
\label{eq:q_diagnostics}
\end{equation}
Values $Q_{\rm tot}\gg1$ indicate that the resolved velocity gradients
dominate the denominator of the viscosity law, whereas
$Q_{\rm tot}\ll1$ indicates that the cutoff term dominates.

For the strongest-viscosity model, the median value is
$Q_{\rm tot}=0.929$, with $Q_{\rm tot}>1$ in
$48.3\%$ of the radial cells.
Thus, the cutoff contribution is globally somewhat
larger than the combined resolved velocity-gradient contribution,
although both remain dynamically relevant.

To quantify the individual contributions, we define the pointwise
fractions
\begin{equation}
f_r
=
\frac{S_r^2}{S_{\rm eff}^2},
\qquad
f_\omega
=
\frac{S_\omega^2}{S_{\rm eff}^2},
\qquad
f_{\min}
=
\frac{S_{\min}^2}{S_{\rm eff}^2}.
\label{eq:seff_fractions}
\end{equation}
By construction, $f_r+f_\omega+f_{\min}=1$ for every radial cell.
For $\alpha_{\rm vis}=1$, the separately computed median contributions
are
${\rm med}(f_r)=0.089$,
${\rm med}(f_\omega)=0.360$, and
${\rm med}(f_{\min})=0.537$.
Because these medians are evaluated separately from the radial
distributions, they are not required to sum exactly to unity even
though the pointwise fractions do.

The radial velocity-gradient contribution therefore does not dominate
$S_{\rm eff}$. Instead, the mixing-length cutoff provides the largest
global contribution, followed by rotational shear, while the radial
velocity-gradient term remains subdominant.

A cell-by-cell comparison leads to the same conclusion. For
$\alpha_{\rm vis}=1$, $S_{\min}$ is the largest contribution in
$54.3\%$ of the radial cells, while rotational shear
dominates in $45.7\%$.
The radial velocity-gradient term is never the largest
contribution. Thus, the adopted viscosity prescription is not
primarily controlled by $dv_r/dr$.

The radial evolution of the $\alpha_{\rm vis}=1$ model can be
quantified over three representative radial intervals.
In the inner wind,
$1\leq r/R_\star<5$, the resolved velocity gradients dominate, with
$Q_{\rm tot}=4.01$, and rotational shear provides the largest
contribution, with $f_\omega=0.859$. Over
$5\leq r/R_\star<50$, the median contribution from the cutoff becomes
the largest, with $f_{\min}=0.7$ and
$Q_{\rm tot}=0.654$. Farther out,
over $50\leq r/R_\star<95$, the cutoff becomes even more dominant, with
$f_{\min}=0.868$, while the rotational contribution decreases to
$f_\omega=0.013$ and the radial velocity-gradient contribution
increases to $f_r=0.119$.

The upper-bound character of the cutoff can also be expressed as
$\nu/(\alpha_{\rm vis}c_s l_{\max})
=S_{\min}/S_{\rm eff}=\sqrt{f_{\min}}$.
For $\alpha_{\rm vis}=1$, the median value of this ratio is
$0.733$.
The effective viscosity therefore remains below the limiting value
$\alpha_{\rm vis}c_s l_{\max}$, while remaining of the same order.

Overall, the adopted prescription operates in a radially varying
regime. Rotational shear dominates the strain-limiting rate in the inner
wind, whereas the mixing-length cutoff controls the intermediate and
outer wind and provides the largest global contribution. The radial
velocity-gradient term becomes more important toward large radii but
never dominates the strain-limiting rate. The cutoff therefore limits
the maximum mixing length and caps the kinematic viscosity at
$\alpha_{\rm vis}c_s l_{\max}$.

The effective viscosity is therefore locally coupled to the resolved
velocity structure while remaining bounded by the prescribed maximum
mixing length. This combination allows the strain-limited prescription
to produce a dynamically relevant radial viscous braking without
driving the wind into a regime dominated by viscous angular-momentum
transport.

The sensitivity of the strongest-viscosity model to the adopted maximum
mixing length is examined in Appendix~\ref{app:lmax_sensitivity}.
Increasing $l_{\max}$ from $0.25r$ to $r$ raises the median kinematic
viscosity by a factor of approximately $2.38$, while
reducing the terminal velocity by approximately $9.2\%$
and the mass-loss rate by approximately $45.5\%$.

\subsection{Radial acceleration decomposition}

\begin{figure}
\centering
\includegraphics[width=\hsize]{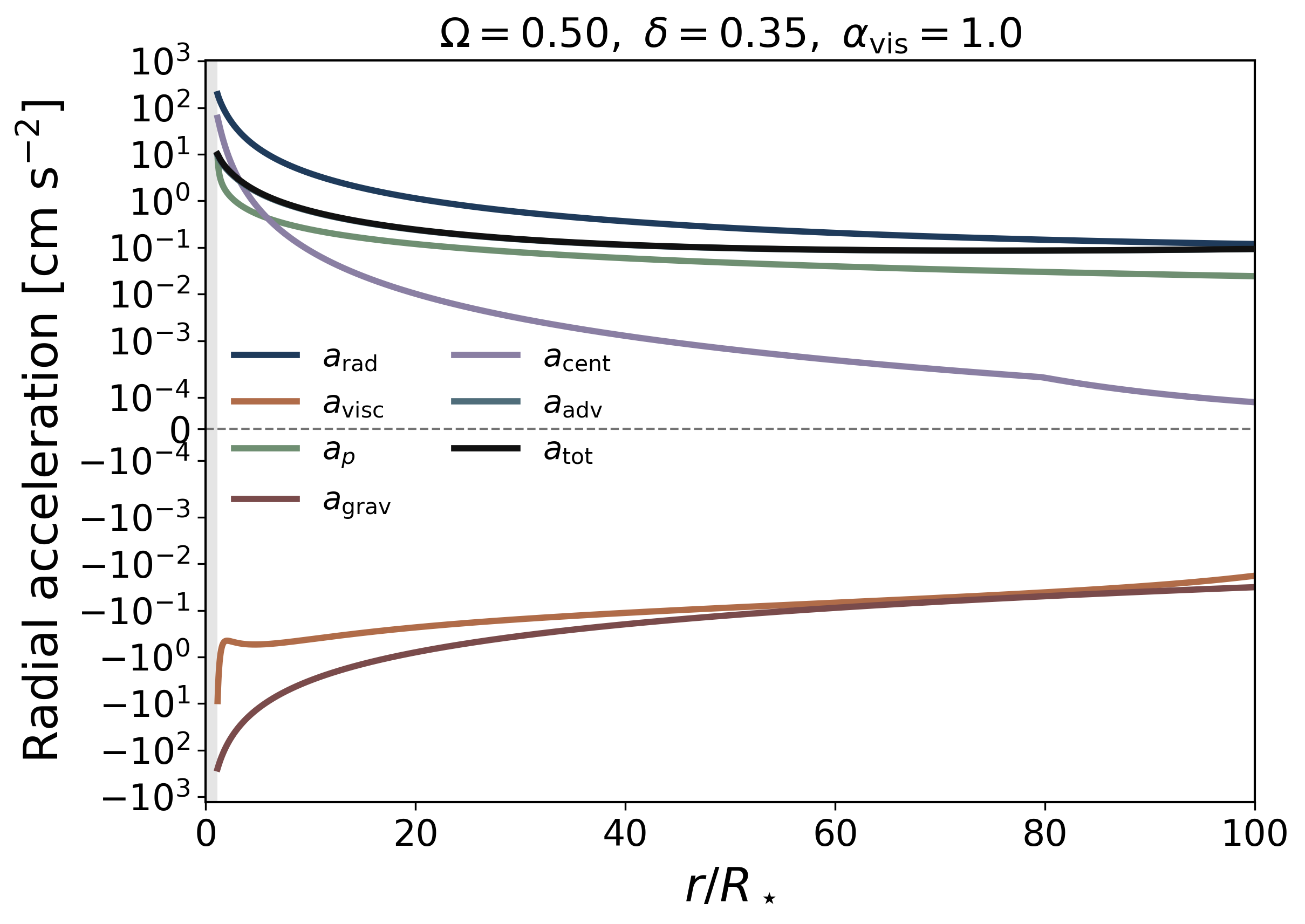}
\caption{
Signed radial acceleration balance for the strongest-viscosity model,
$\alpha_{\rm vis}=1$.
Shown are the total radiative acceleration
$a_{\rm rad}=g_{\rm cont}+g_{\rm line}$, the radial viscous acceleration
$a_{{\rm visc},r}$, the pressure-gradient acceleration $a_p$, the
Newtonian gravitational acceleration $a_{\rm grav}$, the centrifugal
acceleration $a_{\rm cent}$, the advective acceleration
$a_{\rm adv}=v_r\,\partial v_r/\partial r$, and
the total force per unit mass
$a_{\rm tot}=a_{\rm rad}+a_{{\rm visc},r}+a_p+a_{\rm grav}+a_{\rm cent}$.
The grey shaded region marks the inner boundary zone excluded
from quantitative analysis.}
\label{fig:acc_rad_alpha}
\end{figure}

Figure~\ref{fig:acc_rad_alpha} shows the radial acceleration balance
for the strongest-viscosity model, $\alpha_{\rm vis}=1$. The dominant
terms are the outward total radiative acceleration and the inward
Newtonian gravitational acceleration. The pressure-gradient and
centrifugal accelerations remain positive throughout the radial domain
but are subdominant.

The radial viscous acceleration remains negative throughout the radial
domain, confirming that viscosity acts as a systematic radial braking
term. As shown in Fig.~\ref{fig:acc_alpha}, the viscous contribution
remains subdominant to the total radiative acceleration while being
dynamically relevant relative to the advective acceleration. The flow
therefore remains radiatively driven.

For the stationary relaxed solution, $a_{\rm tot}$ closely
follows $a_{\rm adv}$, with a median fractional residual
$|a_{\rm tot}-a_{\rm adv}|/|a_{\rm adv}|=2.53\times10^{-2}$.
Thus, viscosity reduces the net outward acceleration without reversing
the flow, shifting the relaxed solution toward the lower velocities
and mass-loss rates shown in Fig.~\ref{fig:alpha_vel}.

\subsection{Angular-momentum transport}

\begin{figure}
\centering
\includegraphics[width=\hsize]{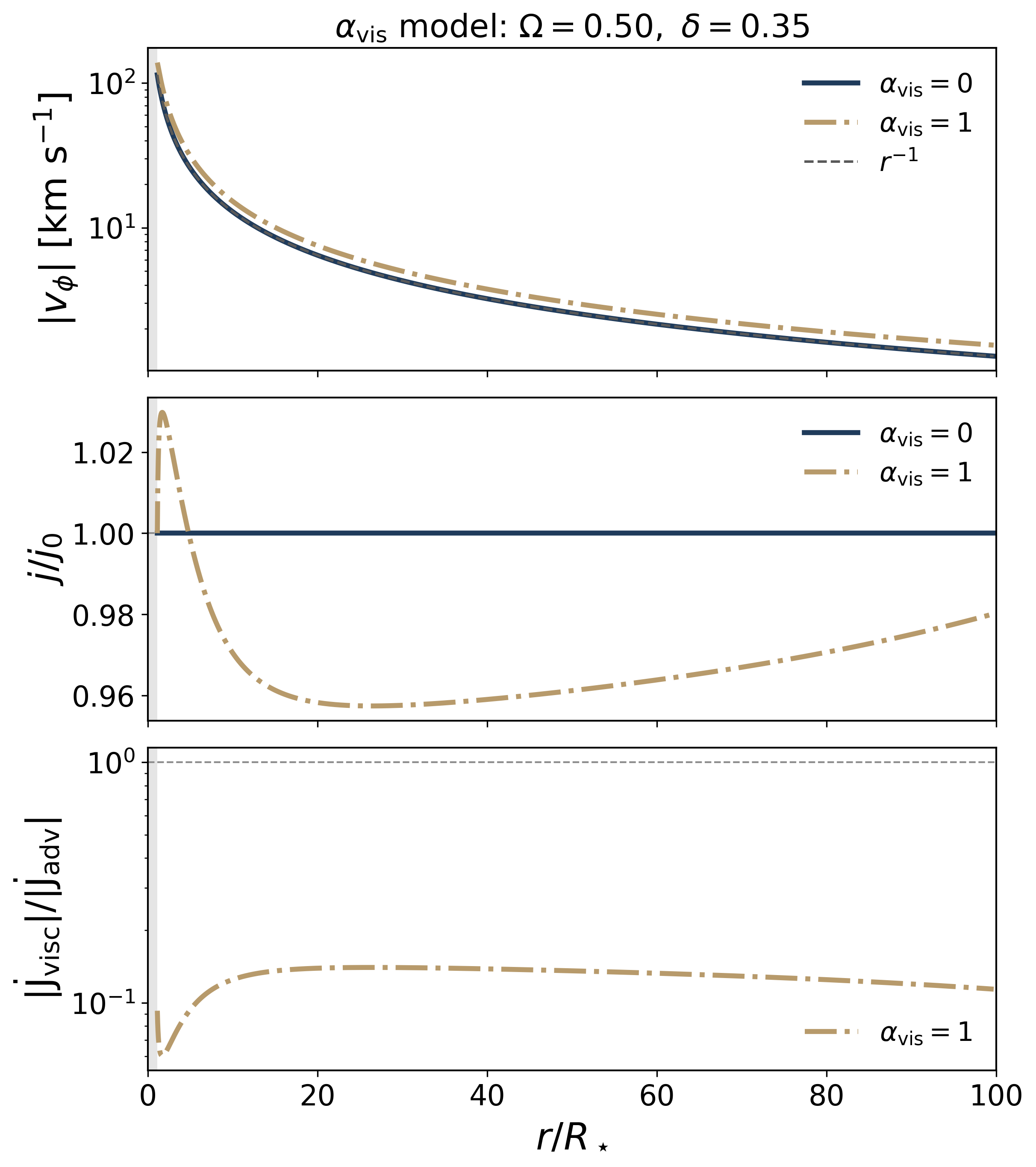}
\caption{
Azimuthal kinematics and angular-momentum transport for the inviscid
and strongest-viscosity models.
Top: azimuthal velocity magnitude $|v_\phi|$, compared with an
$r^{-1}$ reference profile.
Middle: specific angular momentum $j=rv_\phi$, normalized by its inner
reference value $j_0$.
Bottom: ratio of viscous to advective angular-momentum flux for the
$\alpha_{\rm vis}=1$ model. The grey shaded region marks the inner boundary zone excluded
from quantitative analysis.}
\label{fig:angular_momentum_simplified}
\end{figure}

To determine whether viscosity produces significant angular-momentum
redistribution, we consider the specific angular momentum $j=rv_\phi$.
In the absence of viscous transport, $j$ is conserved and therefore
$v_\phi\propto r^{-1}$.

Figure~\ref{fig:angular_momentum_simplified} shows that this scaling is
largely preserved. The inviscid model follows the $r^{-1}$ reference
profile closely, while the $\alpha_{\rm vis}=1$ model shows a modest
departure from angular-momentum conservation, with a median
$j/j_0=0.9733$.

With the spherical formulation,
the advective and viscous angular-momentum fluxes are
\begin{equation}
\begin{aligned}
\dot{J}_{\rm adv}
&=
4\pi r^2\rho\,v_r j,
\\
\dot{J}_{\rm visc}
&=
-4\pi r^4\rho\,\nu
\frac{d\omega}{dr},
\end{aligned}
\label{eq:jdot_fluxes}
\end{equation}
where $\omega=v_\phi/r$ is the local angular velocity.

For the $\alpha_{\rm vis}=1$ model, the viscous flux remains below the
advective flux throughout the radial domain, with a median
$|\dot{J}_{\rm visc}|/|\dot{J}_{\rm adv}|=1.22\times10^{-1}$.
Angular momentum is therefore transported predominantly by advection,
even in the strongest-viscosity model.

The principal dynamical effect of the strain-limited viscosity is thus
radial: it modifies the radial momentum balance and the resulting
mass-loss rate without driving the wind into a regime dominated by
viscous angular momentum redistribution.

\subsection{Wind response to constant viscosity}
\label{sec:constant_nu_response}

We now examine the models with spatially uniform kinematic viscosity.
The adopted values are $\nu=0$, $3.5\times10^{-3}$, and
$7.3\times10^{-3}$ in code units, with the two nonzero values
approximately matching the launch-region effective viscosities
of the $\alpha_{\rm vis}=0.5$ and $\alpha_{\rm vis}=1$ strain-limited
models, respectively.

\begin{figure}
\centering
\includegraphics[width=\hsize]{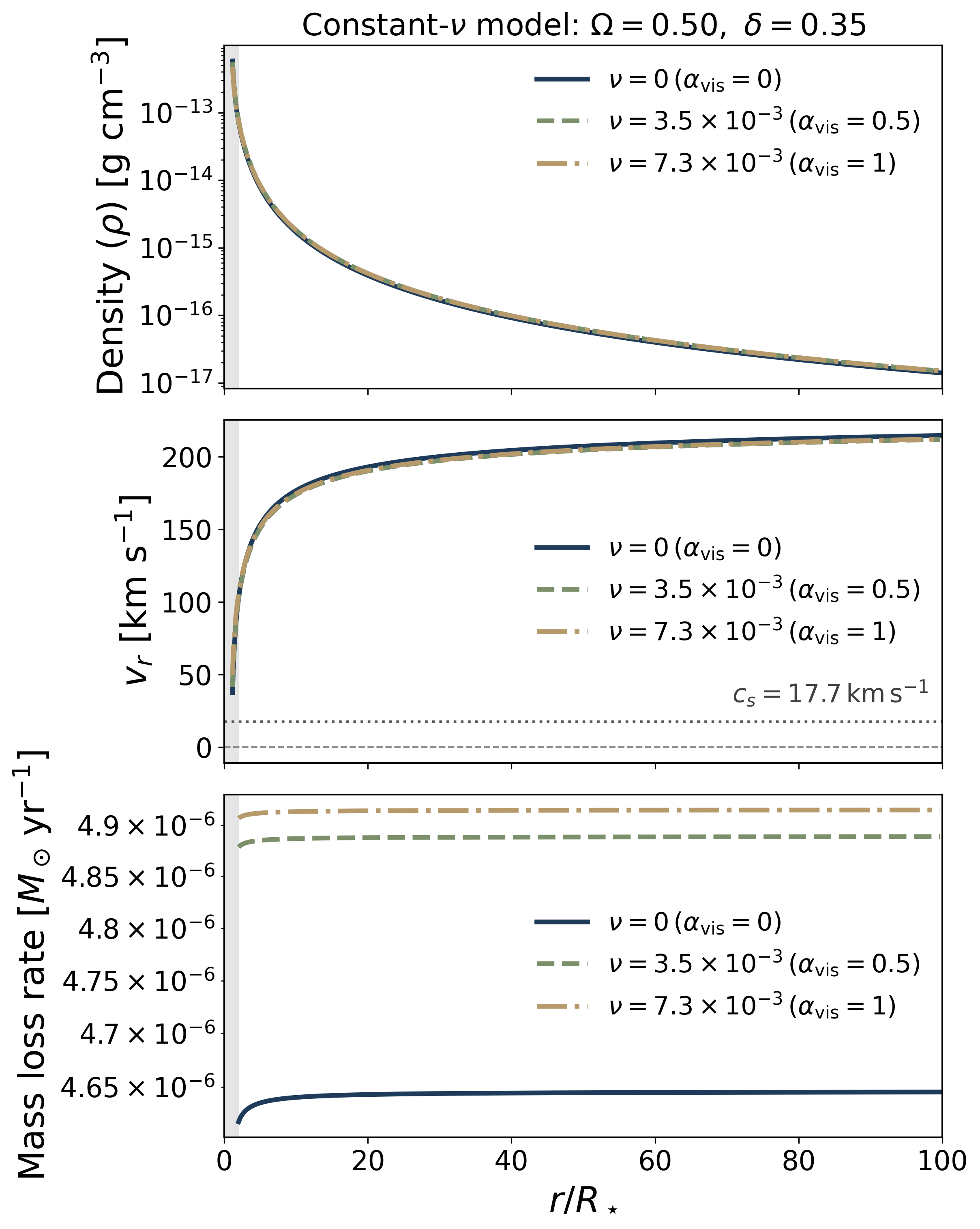}
\caption{
Relaxed radial structure of the constant-viscosity models for
$\Omega=0.50$ and $\delta=0.35$.
The models use $\nu=0$, $3.5\times10^{-3}$, and
$7.3\times10^{-3}$ in code units.
Top: volumetric density $\rho(r)$.
Middle: radial velocity $v_r(r)$. The horizontal dotted line marks
the isothermal sound speed, $c_s=17.7~{\rm km~s^{-1}}$.
Bottom: mass-loss rate $\dot{M}(r)$.
The grey shaded region marks the inner region excluded
from quantitative analysis.}
\label{fig:nu_vel}
\end{figure}

Figure~\ref{fig:nu_vel} shows that increasing the imposed viscosity
produces slightly higher densities and mass-loss rates.
Relative to the inviscid model, the median $\rho$
increases by approximately $6.9\%$ and
$7.2\%$ for
$\nu=3.5\times10^{-3}$ and
$7.3\times10^{-3}$, respectively.
The mass-loss rate increases by approximately
$5.3\%$ and $5.9\%$.

As in the strain-limited models, the radial velocities are
lower than in the inviscid reference, although the changes are much
smaller for constant $\nu$. The median $v_r$ decreases from about
$185.7\,{\rm km\,s^{-1}}$ in the inviscid case to
$182.9$ and $183.4\,{\rm km\,s^{-1}}$ for the two nonzero viscosities,
respectively, corresponding to reductions of approximately
$1.5\%$ and $1.2\%$.

This behavior is consistent with the mass-flux
relation given by Eq.~(\ref{eq:mass_loss_rate}). At fixed radius,
$v_r\propto\dot{M}/\rho$.
The modest density enhancement more than compensates for
the small reduction in radial velocity, yielding a slightly higher
mass-loss rate.

Both viscosity prescriptions produce lower radial velocities
than the inviscid reference, revealing a consistent qualitative
kinematic response. However, the constant-$\nu$ models remain close
to the inviscid solution, whereas the strain-limited models exhibit
substantial reductions in wind velocity and mass-loss rate.
This difference is consistent with the radial increase of the
strain-limited viscosity: matching its magnitude near the launch
region does not reproduce its strength throughout the wind.
The shared decrease in velocity is accompanied by different
density responses, yielding a slight increase in mass-loss rate
for constant $\nu$ and a substantial decrease for the strain-limited
prescription.
This comparison shows that the wind response depends not only on the
magnitude of viscosity, but also on its radial structure.

\subsection{Time evolution of terminal velocity and mass-loss rate}

\begin{figure}
\centering
\includegraphics[width=\columnwidth]{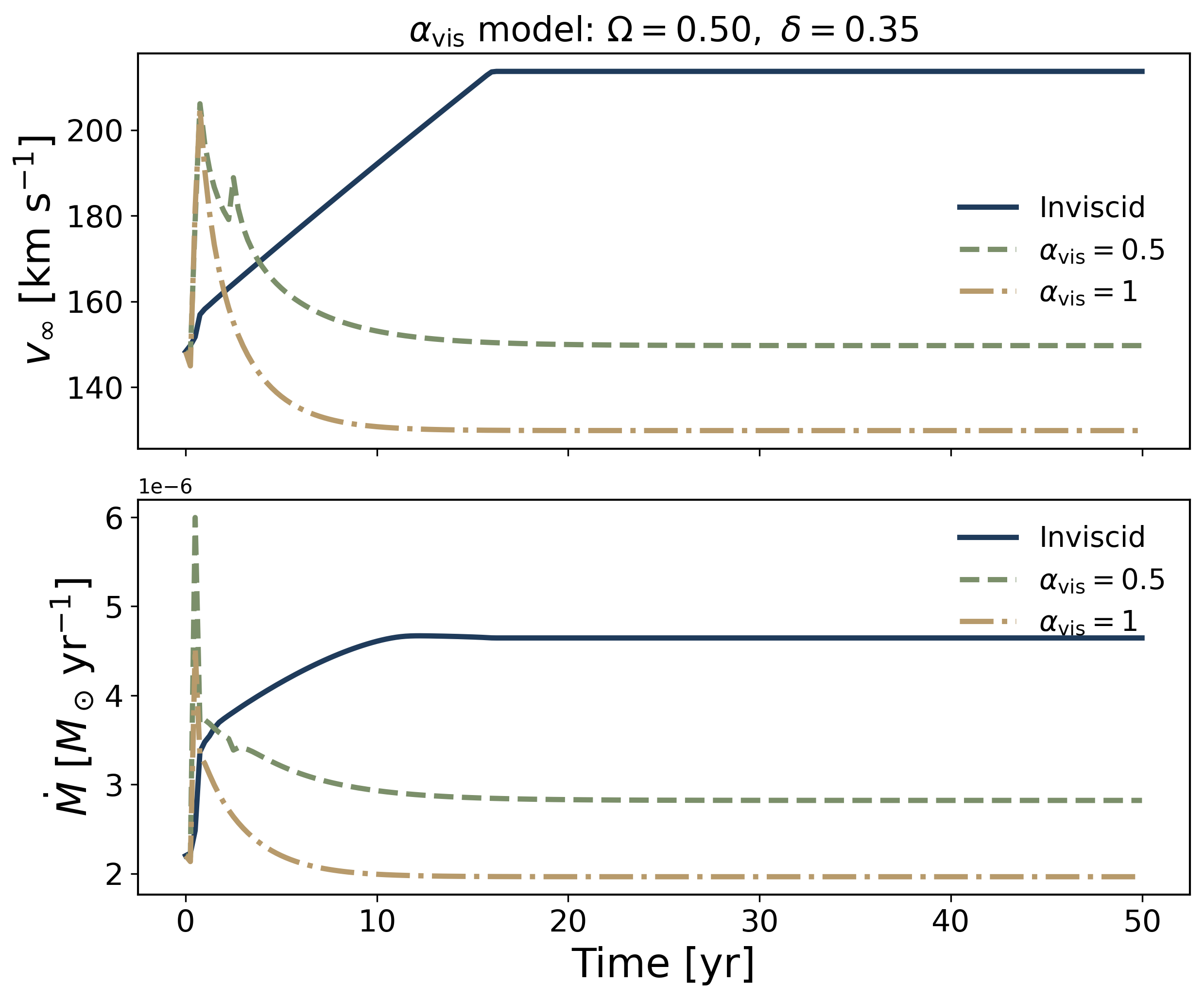}
\caption{
Time evolution of the terminal velocity and mass-loss rate
for the strain-limited $\alpha_{\rm vis}$ models with
$\Omega=0.50$ and $\delta=0.35$.
The models reach a stationary state after approximately
16 yr and remain stationary until the end of the 50 yr integrations.
Increasing $\alpha_{\rm vis}$ lowers both quantities after the initial
transient.
}
\label{fig:loss_rate_evol_alpha}
\end{figure}

Figure~\ref{fig:loss_rate_evol_alpha} shows the time evolution of the
terminal velocity and mass-loss rate for the strain-limited
$\alpha_{\rm vis}$ models over 50 yr.
Following the initial transient, both quantities settle
to stationary values that persist for the remainder of the
simulation. The final outputs therefore characterize fully
relaxed wind solutions.

In the stationary solutions, increasing $\alpha_{\rm vis}$
lowers both $v_\infty$ and $\dot{M}$.
At the final output, the inviscid model reaches
$v_\infty=214.7~{\rm km~s^{-1}}$ and
$\dot{M}=4.64\times10^{-6}~M_\odot~{\rm yr^{-1}}$,
whereas the $\alpha_{\rm vis}=1$ model reaches
$v_\infty=131.8~{\rm km~s^{-1}}$ and
$\dot{M}=1.96\times10^{-6}~M_\odot~{\rm yr^{-1}}$.
Thus, the strongest strain-limited model reduces the terminal velocity
by $38.6\%$ and the mass-loss rate by
$57.7\%$ relative to the inviscid solution.
This temporal evolution is consistent with the radial viscous
braking, as discussed in the previous sections.

\begin{figure}
\includegraphics[width=\columnwidth]{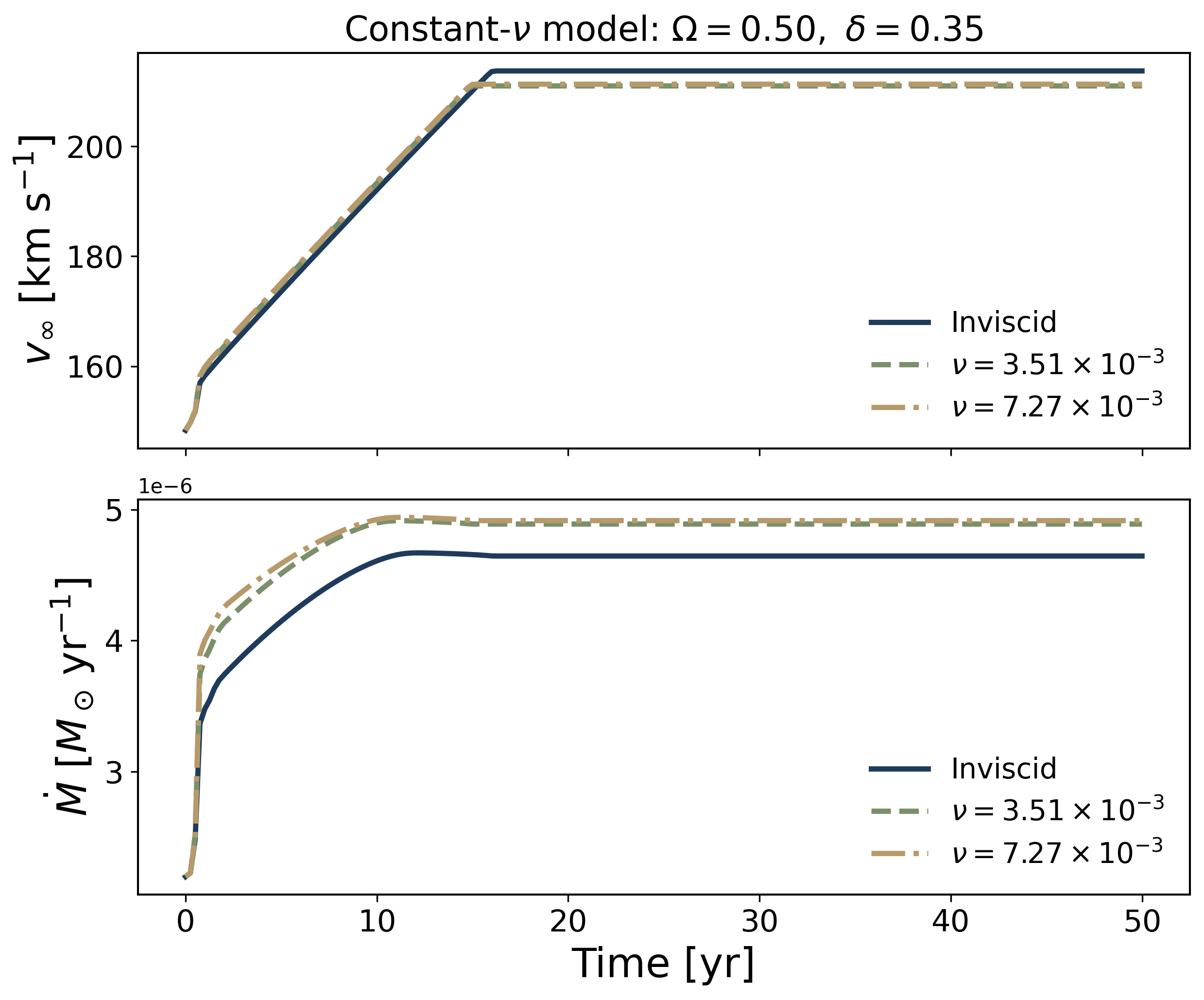}
\caption{
Time evolution of the terminal velocity and mass-loss rate
for the constant-viscosity models with
$\Omega=0.50$ and $\delta=0.35$.
The inviscid reference is compared with the two nonzero viscosities
approximately matched to the launch-region values of the
$\alpha_{\rm vis}=0.5$ and $\alpha_{\rm vis}=1$ models, respectively.
The models reach a stationary state after approximately
16 yr and remain stationary until the end of the 50 yr integrations.
The stationary nonzero-viscosity solutions have slightly higher
mass-loss rates and slightly lower outer wind velocities than
the inviscid reference.}
\label{fig:loss_rate_evol_nu}
\end{figure}

Figure~\ref{fig:loss_rate_evol_nu} shows the corresponding evolution
for the constant-viscosity models.
Both the terminal velocity and mass-loss rate stabilize
after approximately 16 yr and remain stationary until the end
of the 50 yr integrations. The nonzero-viscosity models reach
slightly higher mass-loss rates and slightly lower outer wind
velocities than the inviscid reference.
At the final output, the largest constant viscosity,
$\nu=7.27\times10^{-3}$, gives
$\dot{M}=4.91\times10^{-6}~M_\odot~{\rm yr^{-1}}$ and
$v_\infty=212.3~{\rm km~s^{-1}}$, corresponding to changes
of $+5.9\%$ and $-1.1\%$, respectively,
relative to the inviscid model.

As discussed in Sect.~\ref{sec:constant_nu_response},
the constant-$\nu$ models relax to slightly denser wind
solutions with modestly reduced radial velocities.
The velocity response has the same sign as in the strain-limited
models, but is much weaker, while the mass-loss response has
the opposite sign.

The final wind properties and their relative changes
are summarized in Table~\ref{tab:summary_relaxed_models}.

\begin{table*}
\caption{Summary of the relaxed wind models.}
\label{tab:summary_relaxed_models}
\centering
\small
\begin{tabular}{llcccc}
\hline\hline
Prescription & Parameter & $v_\infty$ & $\Delta v_\infty$
& $\dot{M}$ & $\Delta\dot{M}$ \\
 & & ${\rm km~s^{-1}}$ & $\%$
 & $10^{-6}\,M_\odot\,{\rm yr^{-1}}$ & $\%$ \\
\hline
Inviscid reference & --
& $214.7$ & 0.0
& $4.64$ & 0.0 \\

Strain-limited & $\alpha_{\rm vis}=0.5$
& $150.8$ & $-29.7$
& $2.82$ & $-39.2$ \\

Strain-limited & $\alpha_{\rm vis}=1.0$
& $131.8$ & $-38.6$
& $1.96$ & $-57.7$ \\

Constant $\nu$ & $\nu=3.51\times10^{-3}$
& $212.0$ & $-1.3$
& $4.89$ & $+5.3$ \\

Constant $\nu$ & $\nu=7.27\times10^{-3}$
& $212.3$ & $-1.1$
& $4.91$ & $+5.9$ \\
\hline
\end{tabular}
\tablecomments{
The terminal velocity and mass-loss rate are measured at the final
available output. Relative variations are computed with respect to the
inviscid reference model. The mass-loss rate is calculated as
$\dot{M}=4\pi r^2\rho v_r$. The constant kinematic viscosities are given in code units.
}
\end{table*}

\section{Discussion}

The results show that a strain-limited viscosity prescription can
produce a measurable dynamical response in a supersonic line-driven
wind. The dominant effect is radial: increasing $\alpha_{\rm vis}$
shifts the wind toward lower radial velocities, densities, and
mass-loss rates. For $\alpha_{\rm vis}=1$, the terminal velocity
decreases by about $39\%$ and the mass-loss rate
by about $58\%$ relative to the inviscid model
in the stationary state.
The radial response is associated with an extended negative viscous
acceleration, which opposes the outward wind acceleration while
remaining subdominant to the total radiative acceleration.

This behavior differs from the transport regime assumed in a
Keplerian disk. In the Shakura--Sunyaev framework, the characteristic
viscosity scales as $\nu\sim\alpha c_s H$, with
$H=c_s/\omega_{\rm K}$ under vertical hydrostatic equilibrium.
The present models instead describe radially accelerating,
non-Keplerian equatorial outflows with no physical scale height. Their angular
momentum transport is dominated by advection rather than viscous
redistribution. The standard disk interpretation of the
Shakura--Sunyaev prescription is therefore not directly applicable to
these winds.

The decomposition of $S_{\rm eff}$ shows that the radial
velocity-gradient term, $dv_r/dr$, does not dominate the adopted
viscosity closure. For $\alpha_{\rm vis}=1$, the separately evaluated
median contributions are $f_r=0.089$,
$f_\omega=0.360$, and $f_{\min}=0.537$.
The maximum-mixing-length cutoff therefore provides the largest
global contribution, followed by rotational shear, while the radial
velocity-gradient term remains subdominant. The cutoff imposes
the upper bound $\nu\leq\alpha_{\rm vis}c_s l_{\max}$, preventing
the effective diffusivity from becoming arbitrarily large when
the resolved velocity gradients are small.

The constant-viscosity models show that the wind response
depends on more than the viscosity amplitude near the launch region.
Although the constant values of $\nu$ are approximately matched
to the launch-region values of the corresponding strain-limited
models, they produce only small departures from the inviscid
solution. Both prescriptions yield lower radial velocities, but
the reduction is much stronger when the viscosity increases outward,
as in the strain-limited models. This behavior supports the importance
of the radial viscosity profile for the global wind response.
The mass-loss rate, however, increases slightly for constant $\nu$
and decreases substantially for the strain-limited prescription.
A reduction in wind velocity therefore does not imply a universal
decrease in mass-loss rate; the accompanying density response
must also be considered.

The angular-momentum analysis further shows that, for the wind models
and parameter range explored here, the strain-limited prescription does
not drive the flow into a disk-like transport regime.
For the strongest strain-limited model, $v_\phi$ remains close to the
angular-momentum-conserving scaling $v_\phi\propto r^{-1}$, while the
median ratio of viscous to advective angular-momentum transport is
$|\dot{J}_{\rm visc}|/|\dot{J}_{\rm adv}|=
1.22\times10^{-1}$.
The radial viscous force can therefore modify the wind dynamics without
making viscous angular momentum transport dominant.

The sensitivity test in Appendix~\ref{app:lmax_sensitivity} shows that
$l_{\max}$ controls the amplitude of the effective viscosity and the
resulting wind response. Across the explored range,
$0.25r\leq l_{\max}\leq r$, increasing $l_{\max}$ produces a larger
effective viscosity together with lower terminal velocities and
mass loss rates while preserving the same qualitative wind response.

The present results apply to the adopted $\delta$-slow
model \citep{cure2011}. The same viscosity prescription can be applied to fast and
$\Omega$-slow solutions, but their different velocity gradients,
rotation, and density profiles will modify the viscous response.
At fixed $c_s$ and $\alpha_{\rm vis}$, steeper radial velocity
gradients increase $S_{\rm eff}$ and reduce the local viscosity.
Likewise, stronger rotational shear can limit the viscosity in
rapidly rotating $\Omega$-slow winds. However, the viscous
acceleration depends on the divergence of the stress, including
the density and viscosity gradients, so its dynamical importance
cannot be inferred from the viscosity amplitude alone. The
magnitude and even the sign of the mass-loss response found here
should therefore not be assumed to hold for other solution
branches without dedicated simulations.

Our calculation should nevertheless be regarded as a controlled
one-dimensional experiment. The hydrodynamics is solved in a one-dimensional spherical
volume-density formulation using \texttt{FARGO3D}, with no resolved
meridional structure. Moreover,
$S_{\rm eff}$ is a phenomenological scalar measure constructed from
the radial velocity gradient, rotational shear, and the prescribed
mixing-length cutoff; it is not the full invariant of the spherical
rate-of-strain tensor. The models are also isothermal and do not include
multidimensional structure, radiative cooling, clumping, or the
line-driven instability. These effects may alter both the local
velocity gradients, and effective turbulent transport.

In the inviscid m-CAK framework, the critical-point
regularity conditions constrain the mass-loss rate. The divergence
of the viscous stress introduces second-order spatial derivatives
into the stationary momentum equation, even for constant viscosity,
so the inviscid critical-point conditions cannot be applied
unchanged. Here, the mass-loss rates are measured from the relaxed
time-dependent solutions. We have not determined the critical-point
structure of the viscous system and therefore do not assign a
critical radius or evaluate the viscosity at such a location.
A dedicated analysis of this structure is beyond the scope of
the present study.

Within these limitations, the simulations demonstrate that macroscopic
viscous stresses can modify the radial momentum balance and mass-loss
rate of a supersonic line-driven wind without changing its fundamentally
radiatively driven character or producing strong viscous
angular momentum redistribution.

\section{Conclusions}

We have investigated the dynamical effect of a strain-limited turbulent
viscosity in a time-dependent m-CAK model of a rotating B-supergiant
wind. The adopted prescription,
$\nu(r)=\alpha_{\rm vis}c_s^2/S_{\rm eff}(r)$,
couples the kinematic viscosity to the local radial velocity gradient,
rotational shear and a maximum mixing length cutoff. Our main
conclusions are as follows.

\begin{enumerate}

\item The strain-limited prescription produces a strongly radially
structured viscosity. The kinematic viscosity is smallest near the wind
base, where the resolved velocity gradients are largest, and generally
increases through the outer wind as the effective strain-limiting rate
decreases.

\item Increasing $\alpha_{\rm vis}$ monotonically shifts the
stationary winds toward lower radial velocities, densities,
and mass-loss rates.
For $\alpha_{\rm vis}=1$, the terminal velocity decreases by about
$39\%$ and the mass-loss rate by about
$58\%$ relative to the inviscid model.

\item The main dynamical effect of viscosity is radial braking. For the
nonzero $\alpha_{\rm vis}$ models, $a_{{\rm visc},r}$ remains negative
throughout the radial domain. Even in the strongest-viscosity model,
the viscous acceleration remains subdominant to the total radiative
acceleration while being dynamically relevant relative to the
advective acceleration. The wind therefore remains fundamentally
radiatively driven.

\item The viscosity prescription operates in a mixed regime rather than
being controlled by the radial velocity gradient alone. For
$\alpha_{\rm vis}=1$, the separately evaluated median contributions are
$f_r=0.089$, $f_\omega=0.360$, and
$f_{\min}=0.537$. The maximum-mixing-length cutoff
therefore provides the largest global contribution, followed by
rotational shear, while the radial velocity-gradient term remains
subdominant. Despite this, angular-momentum transport remains
dominated by advection, with $v_\phi$ remaining close to the
angular-momentum-conserving scaling $v_\phi\propto r^{-1}$.

\item The global wind response depends on the radial structure of the
viscosity, not only on its amplitude near the wind base.
Constant-viscosity models approximately matched to the
launch-region values of the strain-limited solutions produce
slightly denser winds with slightly higher mass-loss rates
and lower radial velocities. Both prescriptions therefore share
the same qualitative velocity response, but the strain-limited
models exhibit much stronger changes and a decrease in mass-loss
rate. This comparison supports the importance of the outward
increase in viscosity for the substantial wind modification
found in the strain-limited models.
Increasing the maximum mixing length from $0.25r$ to $r$ raises the
median kinematic viscosity by a factor of approximately
$2.38$, while reducing the terminal velocity by
approximately $9.2\%$ and the mass-loss rate by
approximately $45.5\%$, preserving the same qualitative
wind response.

\end{enumerate}

These results show that macroscopic turbulent viscosity can modify the
radial momentum balance and mass-loss rate of a supersonic line-driven
wind without driving the flow into a disk-like regime dominated by
viscous angular momentum redistribution. Within the present
one-dimensional framework, the dominant effect of
strain-limited viscosity is a dynamically relevant but
subdominant braking of the radial outflow.

\begin{acknowledgements}
We thank the anonymous referee for constructive comments that improved this manuscript.
M.M. and M.C. acknowledge financial support from FONDECYT Regular project 1241818. M.C. and I.A. thank FONDECYT projects 1230131 and 1261498 for their support of this study. MC acknowledges partial support from the Centro de Astrofísica de Valparaíso. This project has been partially co-funded by the European Union under Project 101183150, OCEANS. R.O.J.V. acknowledges financial support from Universidad Nacional de La Plata (Programa de Incentivos 11/G192 and 11/G193).
\end{acknowledgements}


\bibliography{biblio}{} 

@ARTICLE{Cure+2026,
       author = {{Cure}, Michel and {Araya}, Ignacio and {Meneses}, Rodrigo and {Montesinos}, Matias and {Venero}, Roberto O.~J. and {Arcos}, Catalina and {Rodriguez}, Abigali and {Cidale}, Lydia S.},
        title = "{Quasi-Keplerian Be-Star Disks with Mimicking Viscosity}",
      journal = {arXiv e-prints},
         year = 2026,
        month = jul,
          eid = {arXiv:2607.28870},
        pages = {arXiv:2607.28870},
          doi = {10.48550/arXiv.2607.28870},
archivePrefix = {arXiv},
       eprint = {2607.28870},
 primaryClass = {astro-ph.SR},
       adsurl = {https://ui.adsabs.harvard.edu/abs/2026arXiv260728870C}
}

@ARTICLE{Owocki1988,
       author = {{Owocki}, Stanley P. and {Castor}, John I. and {Rybicki}, George B.},
        title = "{Time-dependent Models of Radiatively Driven Stellar Winds. I. Nonlinear Evolution of Instabilities for a Pure Absorption Model}",
      journal = {\apj},
         year = 1988,
        month = dec,
       volume = {335},
        pages = {914},
          doi = {10.1086/166977},
       adsurl = {https://ui.adsabs.harvard.edu/abs/1988ApJ...335..914O}
}

@ARTICLE{Owocki1984,
       author = {{Owocki}, S.~P. and {Rybicki}, G.~B.},
        title = "{Instabilities in line-driven stellar winds. I. Dependence on perturbation wavelength.}",
      journal = {\apj},
         year = 1984,
        month = sep,
       volume = {284},
        pages = {337-350},
          doi = {10.1086/162412},
       adsurl = {https://ui.adsabs.harvard.edu/abs/1984ApJ...284..337O}
}

@ARTICLE{Kudritzki2000,
       author = {{Kudritzki}, Rolf-Peter and {Puls}, Joachim},
        title = "{Winds from Hot Stars}",
      journal = {\araa},
         year = 2000,
        month = jan,
       volume = {38},
        pages = {613-666},
          doi = {10.1146/annurev.astro.38.1.613},
       adsurl = {https://ui.adsabs.harvard.edu/abs/2000ARA&A..38..613K}
}

@ARTICLE{DeFreitasPacheco1976,
       author = {{de Freitas Pacheco}, J.~A. and {Steiner}, J.~E.},
        title = "{The spectrum of Cyg X-1 {\textemdash} A theoretical model}",
      journal = {\apss},
         year = 1976,
        month = feb,
       volume = {39},
       number = {2},
        pages = {487-494},
          doi = {10.1007/BF00648346},
       adsurl = {https://ui.adsabs.harvard.edu/abs/1976Ap&SS..39..487D}
}

@ARTICLE{Benitez-Llambay+2016,
       author = {{Ben{\'\i}tez-Llambay}, Pablo and {Masset}, Fr{\'e}d{\'e}ric S.},
        title = "{FARGO3D: A New GPU-oriented MHD Code}",
      journal = {\apjs},
         year = 2016,
        month = mar,
       volume = {223},
       number = {1},
          eid = {11},
        pages = {11},
          doi = {10.3847/0067-0049/223/1/11},
archivePrefix = {arXiv},
       eprint = {1602.02359},
 primaryClass = {astro-ph.IM},
       adsurl = {https://ui.adsabs.harvard.edu/abs/2016ApJS..223...11B}
}

@ARTICLE{Masset2000,
       author = {{Masset}, F.},
        title = "{FARGO: A fast eulerian transport algorithm for differentially rotating disks}",
      journal = {\aaps},
         year = 2000,
        month = jan,
       volume = {141},
        pages = {165-173},
          doi = {10.1051/aas:2000116},
archivePrefix = {arXiv},
       eprint = {astro-ph/9910390},
 primaryClass = {astro-ph},
       adsurl = {https://ui.adsabs.harvard.edu/abs/2000A&AS..141..165M}
}

@BOOK{Landau1987,
       author = {{Landau}, L.~D. and {Lifshitz}, E.~M.},
        title = "{Fluid Mechanics}",
         year = 1987,
       adsurl = {https://ui.adsabs.harvard.edu/abs/1987flme.book.....L}
}

@ARTICLE{Jiang2015,
       author = {{Jiang}, Yan-Fei and {Cantiello}, Matteo and {Bildsten}, Lars and {Quataert}, Eliot and {Blaes}, Omer},
        title = "{Local Radiation Hydrodynamic Simulations of Massive Star Envelopes at the Iron Opacity Peak}",
      journal = {\apj},
         year = 2015,
        month = nov,
       volume = {813},
       number = {1},
          eid = {74},
        pages = {74},
          doi = {10.1088/0004-637X/813/1/74},
archivePrefix = {arXiv},
       eprint = {1509.05417},
 primaryClass = {astro-ph.SR},
       adsurl = {https://ui.adsabs.harvard.edu/abs/2015ApJ...813...74J}
}

@ARTICLE{Debnath2024,
       author = {{Debnath}, D. and {Sundqvist}, J.~O. and {Moens}, N. and {Van der Sijpt}, C. and {Verhamme}, O. and {Poniatowski}, L.~G.},
        title = "{2D unified atmosphere and wind simulations of O-type stars}",
      journal = {\aap},
         year = 2024,
        month = apr,
       volume = {684},
          eid = {A177},
        pages = {A177},
          doi = {10.1051/0004-6361/202348206},
archivePrefix = {arXiv},
       eprint = {2401.08391},
 primaryClass = {astro-ph.SR},
       adsurl = {https://ui.adsabs.harvard.edu/abs/2024A&A...684A.177D}
}

@ARTICLE{okazaki2001,
   author = {{Okazaki}, A.~T.},
    title = "{Viscous Transonic Decretion in Disks of Be Stars}",
  journal = {Publications of the Astronomical Society of Japan},
   eprint = {astro-ph/0010517},
     year = 2001,
    month = feb,
   volume = 53,
    pages = {119-125},
      doi = {10.1093/pasj/53.1.119},
   adsurl = {http://adsabs.harvard.edu/abs/2001PASJ...53..119O}
}

@ARTICLE{lee1991,
   author = {{Lee}, U. and {Osaki}, Y. and {Saio}, H.},
    title = "{Viscous excretion discs around Be stars}",
  journal = {Monthly Notices of the Royal Astronomical Society},
     year = 1991,
    month = may,
   volume = 250,
    pages = {432-437},
      doi = {10.1093/mnras/250.2.432},
   adsurl = {http://adsabs.harvard.edu/abs/1991MNRAS.250..432L}
}

@ARTICLE{owocki1996,
   author = {{Owocki}, S.~P. and {Cranmer}, S.~R. and {Gayley}, K.~G.},
    title = "{Inhibition FO Wind Compressed Disk Formation by Nonradial Line-Forces in Rotating Hot-Star Winds}",
  journal = {The Astrophysical Journal},
     year = 1996,
    month = dec,
   volume = 472,
    pages = {L115},
      doi = {10.1086/310372},
   adsurl = {http://adsabs.harvard.edu/abs/1996ApJ...472L.115O}
}

@ARTICLE{bjorkman1993,
   author = {{Bjorkman}, J.~E. and {Cassinelli}, J.~P.},
    title = "{Equatorial disk formation around rotating stars due to Ram pressure confinement by the stellar wind}",
  journal = {The Astrophysical Journal},
     year = 1993,
    month = may,
   volume = 409,
    pages = {429-449},
      doi = {10.1086/172676},
   adsurl = {http://adsabs.harvard.edu/abs/1993ApJ...409..429B}
}

@ARTICLE{pelupessy2000,
   author = {{Pelupessy}, I. and {Lamers}, H.~J.~G.~L.~M. and {Vink}, J.~S.
	},
    title = "{The radiation driven winds of rotating B[e] supergiants}",
  journal = {Astronomy \&  Astrophysics},
   eprint = {astro-ph/0005300},
     year = 2000,
    month = jul,
   volume = 359,
    pages = {695-706},
   adsurl = {http://adsabs.harvard.edu/abs/2000A%26A...359..695P}
}

@ARTICLE{dessart2005,
   author = {{Dessart}, L. and {Owocki}, S.~P.},
    title = "{2D simulations of the line-driven instability in hot-star winds. II. Approximations for the 2D radiation force}",
  journal = {Astronomy \&  Astrophysics},
   eprint = {astro-ph/0503514},
     year = 2005,
    month = jul,
   volume = 437,
    pages = {657-666},
      doi = {10.1051/0004-6361:20052778},
   adsurl = {http://adsabs.harvard.edu/abs/2005A%26A...437..657D}
}

@ARTICLE{cure2011,
   author = {{Cur{\'e}}, M. and {Cidale}, L. and {Granada}, A.},
    title = "{Slow Radiation-driven Wind Solutions of A-type Supergiants}",
  journal = {The Astrophysical Journal},
archivePrefix = "arXiv",
   eprint = {1105.5576},
 primaryClass = "astro-ph.SR",
     year = 2011,
    month = aug,
   volume = 737,
      eid = {18},
    pages = {18},
      doi = {10.1088/0004-637X/737/1/18},
   adsurl = {http://adsabs.harvard.edu/abs/2011ApJ...737...18C}
}

@ARTICLE{friend1986,
   author = {{Friend}, D.~B. and {Abbott}, D.~C.},
    title = "{The theory of radiatively driven stellar winds. III - Wind models with finite disk correction and rotation}",
  journal = {The Astrophysical Journal},
     year = 1986,
    month = dec,
   volume = 311,
    pages = {701-707},
      doi = {10.1086/164809},
   adsurl = {http://adsabs.harvard.edu/abs/1986ApJ...311..701F}
}

@ARTICLE{puls2008,
       author = {{Puls}, Joachim and {Vink}, Jorick S. and {Najarro}, Francisco},
        title = "{Mass loss from hot massive stars}",
      journal = {\aapr},
         year = 2008,
        month = dec,
       volume = {16},
       number = {3-4},
        pages = {209-325},
          doi = {10.1007/s00159-008-0015-8},
archivePrefix = {arXiv},
       eprint = {0811.0487},
 primaryClass = {astro-ph},
       adsurl = {https://ui.adsabs.harvard.edu/abs/2008A&ARv..16..209P}
}

@ARTICLE{cure2022,
       author = {{Cur{\'e}}, Michel and {Meneses}, Rodrigo and {Araya}, Ignacio and {Arcos}, Catalina and {Pe{\~n}a}, Greco and {Machuca}, Natalia and {Rodriguez}, Abigali},
        title = "{Revisiting viscous transonic decretion disks of Be stars}",
      journal = {Astronomy \&  Astrophysics},
         year = 2022,
        month = aug,
       volume = {664},
          eid = {A185},
        pages = {A185},
          doi = {10.1051/0004-6361/202243620},
archivePrefix = {arXiv},
       eprint = {2206.09031},
 primaryClass = {astro-ph.SR},
       adsurl = {https://ui.adsabs.harvard.edu/abs/2022A&A...664A.185C}
}

@ARTICLE{Machuca2026,
       author = {{Machuca}, N. and {Cur{\'e}}, M. and {Araya}, I. and {Venero}, R.~O.~J. and {Cidale}, L.~S. and {Arcos}, C. and {Sim{\'o}n-D{\'\i}az}, S. and {Lobel}, A.},
        title = "{ISOSCELES Project: II. Modelling galactic B-type stars for fast and $δ-$slow wind regimes}",
      journal = {arXiv e-prints},
         year = 2026,
        month = jun,
          eid = {arXiv:2606.18490},
        pages = {arXiv:2606.18490},
          doi = {10.48550/arXiv.2606.18490},
archivePrefix = {arXiv},
       eprint = {2606.18490},
 primaryClass = {astro-ph.SR},
       adsurl = {https://ui.adsabs.harvard.edu/abs/2026arXiv260618490M}
}

@ARTICLE{Venero2024,
       author = {{Venero}, R.~O.~J. and {Cur{\'e}}, M. and {Puls}, J. and {Cidale}, L.~S. and {Haucke}, M. and {Araya}, I. and {Gormaz-Matamala}, A. and {Arcos}, C.},
        title = "{The wind of rotating B supergiants - II. The {\ensuremath{\delta}}-slow hydrodynamic regime}",
      journal = {\mnras},
         year = 2024,
        month = jan,
       volume = {527},
       number = {1},
        pages = {93-111},
          doi = {10.1093/mnras/stad3030},
       adsurl = {https://ui.adsabs.harvard.edu/abs/2024MNRAS.527...93V}
}

@ARTICLE{Krticka2024,
       author = {{Krti{\v{c}}ka}, J. and {Kub{\'a}t}, J. and {Krti{\v{c}}kov{\'a}}, I.},
        title = "{New mass-loss rates of Magellanic Cloud B supergiants from global wind models}",
      journal = {\aap},
         year = 2024,
        month = jan,
       volume = {681},
          eid = {A29},
        pages = {A29},
          doi = {10.1051/0004-6361/202347916},
archivePrefix = {arXiv},
       eprint = {2311.01257},
 primaryClass = {astro-ph.SR},
       adsurl = {https://ui.adsabs.harvard.edu/abs/2024A&A...681A..29K}
}

@ARTICLE{Krticka2021,
       author = {{Krti{\v{c}}ka}, J. and {Kub{\'a}t}, J. and {Krti{\v{c}}kov{\'a}}, I.},
        title = "{New mass-loss rates of B supergiants from global wind models}",
      journal = {\aap},
         year = 2021,
        month = mar,
       volume = {647},
          eid = {A28},
        pages = {A28},
          doi = {10.1051/0004-6361/202039900},
archivePrefix = {arXiv},
       eprint = {2101.04973},
 primaryClass = {astro-ph.SR},
       adsurl = {https://ui.adsabs.harvard.edu/abs/2021A&A...647A..28K}
}

@ARTICLE{Haucke2018,
       author = {{Haucke}, M. and {Cidale}, L.~S. and {Venero}, R.~O.~J. and {Cur{\'e}}, M. and {Kraus}, M. and {Kanaan}, S. and {Arcos}, C.},
        title = "{Wind properties of variable B supergiants. Evidence of pulsations connected with mass-loss episodes}",
      journal = {\aap},
         year = 2018,
        month = jun,
       volume = {614},
          eid = {A91},
        pages = {A91},
          doi = {10.1051/0004-6361/201731678},
archivePrefix = {arXiv},
       eprint = {1902.01341},
 primaryClass = {astro-ph.SR},
       adsurl = {https://ui.adsabs.harvard.edu/abs/2018A&A...614A..91H}
}

@ARTICLE{Fullerton1997,
       author = {{Fullerton}, A.~W. and {Massa}, D.~L. and {Prinja}, R.~K. and {Owocki}, S.~P. and {Cranmer}, S.~R.},
        title = "{Wind variability of B supergiants. III. Corotating spiral structures in the stellar wind of HD 64760.}",
      journal = {\aap},
         year = 1997,
        month = nov,
       volume = {327},
        pages = {699-720},
       adsurl = {https://ui.adsabs.harvard.edu/abs/1997A&A...327..699F}
}

@ARTICLE{Prinja1995,
       author = {{Prinja}, Raman K. and {Massa}, Derck and {Fullerton}, Alexander W.},
        title = "{The IUE MEGA Campaign: Modulated Structure in the Wind of HD 64760 (B0.5 Ib)}",
      journal = {\apjl},
         year = 1995,
        month = oct,
       volume = {452},
        pages = {L61},
          doi = {10.1086/309708},
       adsurl = {https://ui.adsabs.harvard.edu/abs/1995ApJ...452L..61P}
}

@ARTICLE{Moens2025,
       author = {{Moens}, N. and {Debnath}, D. and {Verhamme}, O. and {Backs}, F. and {Van der Sijpt}, C. and {Sundqvist}, J.~O. and {Sander}, A.~A.~C.},
        title = "{2D unified atmosphere and wind simulations for a grid of O-type stars}",
      journal = {\aap},
         year = 2025,
        month = dec,
       volume = {704},
          eid = {A121},
        pages = {A121},
          doi = {10.1051/0004-6361/202556825},
archivePrefix = {arXiv},
       eprint = {2511.04542},
 primaryClass = {astro-ph.SR},
       adsurl = {https://ui.adsabs.harvard.edu/abs/2025A&A...704A.121M}
}

@ARTICLE{Shakura1973,
       author = {{Shakura}, N.~I. and {Sunyaev}, R.~A.},
        title = "{Black holes in binary systems. Observational appearance.}",
      journal = {\aap},
         year = 1973,
        month = jan,
       volume = {24},
        pages = {337-355},
       adsurl = {https://ui.adsabs.harvard.edu/abs/1973A&A....24..337S}
}

@ARTICLE{cak,
   author = {{Castor}, J.~I. and {Abbott}, D.~C. and {Klein}, R.~I.},
    title = "{Radiation-driven winds in Of stars}",
  journal = {\apj},
     year = 1975,
    month = jan,
   volume = 195,
    pages = {157-174},
      doi = {10.1086/153315},
   adsurl = {http://adsabs.harvard.edu/abs/1975ApJ...195..157C}
}

@ARTICLE{Pauldrach+1986,
       author = {{Pauldrach}, A. and {Puls}, J. and {Kudritzki}, R.~P.},
        title = "{Radiation-driven winds of hot luminous stars. Improvements of the theory and first results.}",
      journal = {\aap},
         year = 1986,
        month = aug,
       volume = {164},
        pages = {86-100},
       adsurl = {https://ui.adsabs.harvard.edu/abs/1986A&A...164...86P}
}

@ARTICLE{Cure_review_2023,
       author = {{Cur{\'e}}, Michel and {Araya}, Ignacio},
        title = "{Radiation-Driven Wind Hydrodynamics of Massive Stars: A Review}",
      journal = {Galaxies},
         year = 2023,
        month = may,
       volume = {11},
       number = {3},
          eid = {68},
        pages = {68},
          doi = {10.3390/galaxies11030068},
archivePrefix = {arXiv},
       eprint = {2305.11666},
 primaryClass = {astro-ph.SR},
       adsurl = {https://ui.adsabs.harvard.edu/abs/2023Galax..11...68C}
}

@ARTICLE{Montesinos+2026,
       author = {{Montesinos}, Mat{\'\i}as and {Zabala-Arroyo}, Emil and {Castro-Salgado}, Juan Jos{\'e} and {Cur{\'e}}, Michel and {Araya}, Ignacio},
        title = "{Line-driven Radiative Winds in B-supergiants: Bridging the Gap between Fast and Slow m-CAK Solutions}",
      journal = {\apj},
         year = 2026,
        month = may,
       volume = {1003},
       number = {1},
          eid = {14},
        pages = {14},
          doi = {10.3847/1538-4357/ae5239},
archivePrefix = {arXiv},
       eprint = {2603.13555},
 primaryClass = {astro-ph.SR},
       adsurl = {https://ui.adsabs.harvard.edu/abs/2026ApJ..1003...14M}
}

@ARTICLE{Montesinos2011,
       author = {{Montesinos}, M. and {de Freitas Pacheco}, J.~A.},
        title = "{The growth of supermassive black holes fed by accretion disks}",
      journal = {\aap},
         year = 2011,
        month = feb,
       volume = {526},
          eid = {A146},
        pages = {A146},
          doi = {10.1051/0004-6361/201015026},
archivePrefix = {arXiv},
       eprint = {1008.4150},
 primaryClass = {astro-ph.CO},
       adsurl = {https://ui.adsabs.harvard.edu/abs/2011A&A...526A.146M}
}

@ARTICLE{Duschl+2000,
       author = {{Duschl}, W.~J. and {Strittmatter}, P.~A. and {Biermann}, P.~L.},
        title = "{A note on hydrodynamic viscosity and selfgravitation in accretion disks}",
      journal = {\aap},
         year = 2000,
        month = may,
       volume = {357},
        pages = {1123-1132},
          doi = {10.48550/arXiv.astro-ph/0004076},
archivePrefix = {arXiv},
       eprint = {astro-ph/0004076},
 primaryClass = {astro-ph},
       adsurl = {https://ui.adsabs.harvard.edu/abs/2000A&A...357.1123D}
}
\bibliographystyle{aasjournalv7}
\begin{appendix}

\section{Sensitivity to the maximum mixing length}
\label{app:lmax_sensitivity}

The fiducial strain-limited prescription adopts $l_{\max}=0.5r$.
Because the cutoff term
\begin{equation}
S_{\min}
=
\frac{c_s}{l_{\max}}
\end{equation}
contributes significantly to $S_{\rm eff}$, we test the sensitivity of
the strongest-viscosity model, $\alpha_{\rm vis}=1$, to the adopted
maximum mixing length. We consider
\begin{equation}
l_{\max}
=
0.25r,\qquad 0.5r,\qquad r.
\end{equation}

Figure~\ref{fig:lmax_sensitivity_profiles} shows the resulting
kinematic viscosity, radial velocity, and mass loss profiles.
Increasing $l_{\max}$ reduces $S_{\min}$ and therefore permits a larger
effective viscosity where the cutoff contributes significantly to
$S_{\rm eff}$. The main properties of the three relaxed solutions are
summarized in Table~\ref{tab:lmax_sensitivity}.

\begin{table}
\centering
\caption{Sensitivity to the maximum mixing length}
\label{tab:lmax_sensitivity}
\begin{tabular}{cccc}
\hline
$l_{\max}/r$
& $\nu_{\rm med}$
& $v_\infty$
& $\dot{M}$ \\
&
$[10^{19}\,{\rm cm^2\,s^{-1}}]$
& $[{\rm km\,s^{-1}}]$
& $[10^{-6}\,M_\odot\,{\rm yr^{-1}}]$ \\
\hline
0.25 & $1.64$ & $143.5$
     & $2.84$ \\
0.50 & $2.83$ & $131.8$
     & $1.96$ \\
1.00 & $3.91$ & $130.3$
     & $1.55$ \\
\hline
\end{tabular}
\end{table}

Across the full range from $l_{\max}=0.25r$ to $r$, the median
kinematic viscosity increases by a factor of approximately
$2.38$, while the terminal velocity and mass-loss rate
decrease by approximately $9.2\%$ and
$45.5\%$, respectively. Thus, increasing the
maximum mixing length systematically shifts the relaxed solution toward
larger effective viscosity, lower terminal velocity, and lower
mass-loss rate.

These tests show that $l_{\max}$ controls the amplitude of the
effective viscosity, and therefore, the strength of the wind response.
Across the explored range, increasing $l_{\max}$ consistently produces
a larger effective viscosity, a lower terminal velocity, and a lower
mass-loss rate while preserving the same qualitative wind response.

\begin{figure}
\centering
\includegraphics[width=0.85\columnwidth]{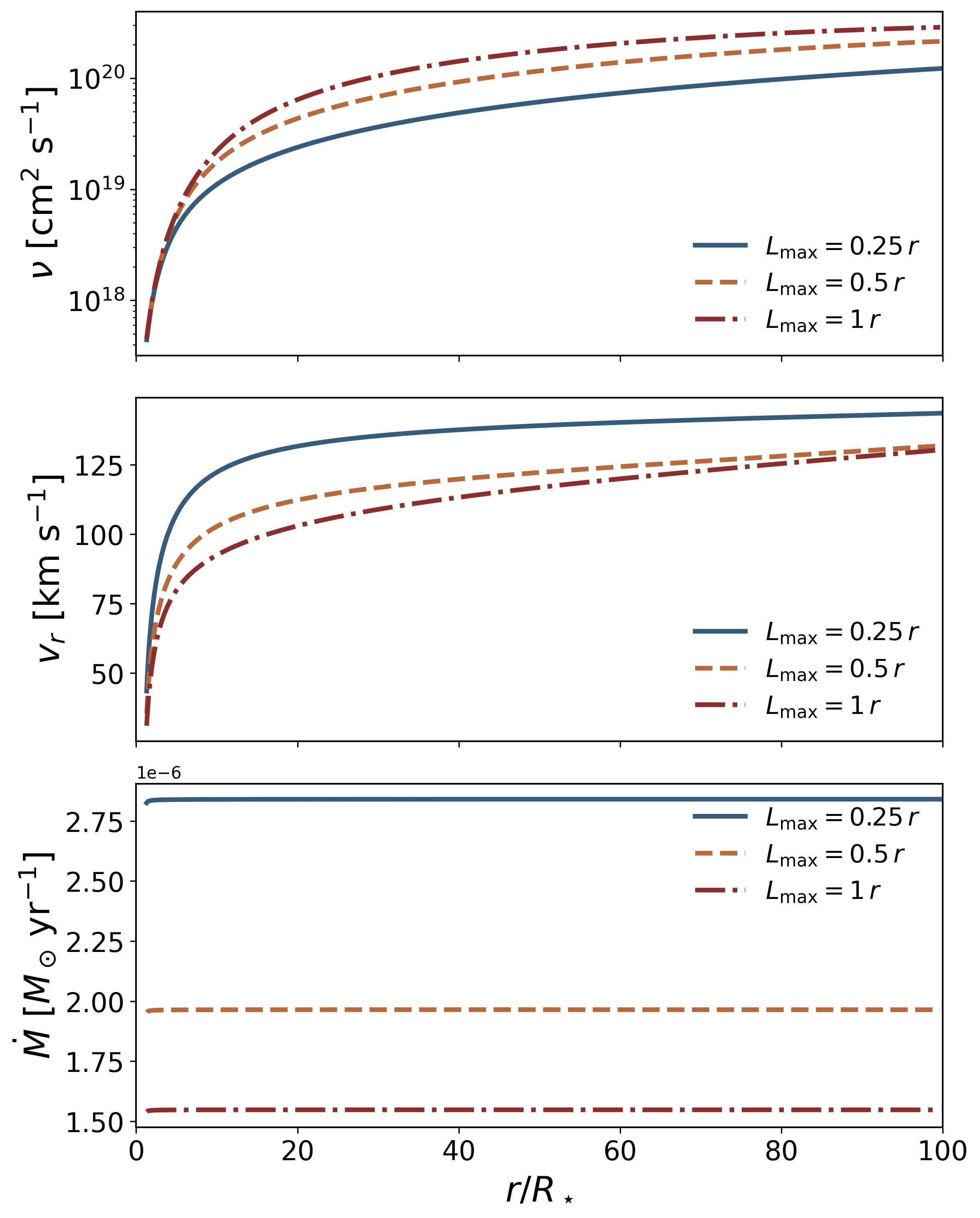}
\caption{
Sensitivity of the $\alpha_{\rm vis}=1$ model to the maximum mixing
length for $\Omega=0.50$ and $\delta=0.35$.
The three cases correspond to $l_{\max}=0.25r$, $0.5r$, and $r$.
Top: Kinematic viscosity $\nu(r)$.
Middle: radial velocity $v_r(r)$.
Bottom: mass-loss rate $\dot{M}(r)$.
Increasing $l_{\max}$ increases the effective viscosity and shifts the
relaxed wind toward lower radial velocities and mass loss rates.
}
\label{fig:lmax_sensitivity_profiles}
\end{figure}

\end{appendix}
\end{document}